\documentclass[twocolumn]{aa}  

\usepackage{natbib}
\usepackage{amsmath}
\usepackage{txfonts}
\usepackage{bm}
\bibpunct{(}{)}{;}{a}{}{,} % to follow the A&A style

\usepackage{color}
\usepackage{graphicx}

\usepackage{txfonts}
\usepackage[draft]{hyperref}
\newcommand{\beq}{\begin{equation}}
\newcommand{\eeq}{\end{equation}}
\newcommand{\bea}{\begin{eqnarray}}
\newcommand{\eea}{\end{eqnarray}}

\newcommand{\cm}{\mathcal}
\newcommand{\mb}{\mathbf}

\newcommand{\kB}{k_\mathrm{B}}

\definecolor{azul}{rgb}{0,0,.8}
\definecolor{rojo}{rgb}{1,0,0}
\begin{document}

\title{Hydrogen ionization equilibrium in magnetic fields}

\author{Mat\'ias Vera Rueda \and  Ren\'e D. Rohrmann} 
 
\institute{ Instituto de Ciencias Astron\'omicas, de la Tierra y del
Espacio (CONICET), Av. Espa\~na 1512 (sur), 5400 San Juan, Argentina}

\abstract{We assess the partition function and ionization degree of magnetized hydrogen atoms at thermodynamic equilibrium for a wide range of field intensities, $B\approx 10^5$~--~$10^{12}$~G.
Evaluations include fitting formulae for an arbitrary number of binding energies, 
the coupling between the internal atomic structure and the center-of-mass motion across the magnetic field, and the formation of the so-called decentered states (bound states with the electron shifted from the Coulomb well). 
Non-ideal gas effects are treated within the occupational probability method.
We also present general mathematical expressions for the bound state correspondence between the limits of zero-field and high-field. This let us evaluate the atomic partition function in a continuous way from the Zeeman perturbative regime to very strong fields. 
Results are shown for conditions found in atmospheres of magnetic white dwarf stars (MWDs), with temperatures $T\approx 5000$~--~$80000$~K and densities $\rho\approx 10^{-12}$~--~$10^{-3}$~g~cm$^3$. 
Our evaluations show a marked reduction of the gas ionization due to the magnetic field in the atmospheres of strong MWDs. We also found that decentered states could be present in the atmospheres of currently known hot MWDs, giving a significant contribution to the partition function in the strongest magnetized atmospheres.
}


\maketitle
%
%________________________________________________________________

%%%%%%%%%%%%%%%%%%%%%%%%%%%%%%%%%%%%%%%%%%%%%%%%%%%%%%%%%%%%%%%%%
\section{Introduction}\label{s:intro}

Strongest magnetic fields are found in stellar objects, compreheding magnetic
white dwarf (MWD) stars with field strengths in a broad range $B\approx 10^3$-$10^9$~G \citep{ferrario:2015} (the lower value likely being a result of detection method limits), neutron stars $10^{8}$-$10^{13}$~G \citep{konar:2017}, and magnetars (i.e., highly magnetized neutron stars)  $10^{14}$-$10^{15}$~G \citep{harding:2006, kaspi:2017}. In comparison, sunspots usually have $B\la 10^3$~G, Ap/Bp-type main-senquence stars $10^3$-$10^4$~G, and the strongest stable magnetic fields generated in terrestrial experiments reach about $10^{5}$~G \citep{crow:1995}, although higher fields ($\approx 10^7$~G) can be produced in magnetic-flux compressions with lifetime very short $\tau\la 10^{-7}$~s \citep{herlach:1999, gotchev:2009}.  Among the mentioned magnetic systems, only magnetars have fields above the critical value $B_c=4.414\times10^{13}$~G where the cyclotron energy exceeds the electron rest energy and particle motions become relativistic ($B_c$ is defined by setting $\hbar\omega_\text{e}=m_\text{e}c^2$, where $\omega_\text{e}=eB/m_\text{e}c$ is the cyclotron frequency and conventional notation is used).

Most compact stars have hydrogen in the outer layers, mainly because the element segregation in their strong gravitational fields. 
The structure of atoms is considerably modified in magnetized compact stars beyond the Zeeman-type perturbative approach. These changes have consequences on the internal partition function, energy level populations, ionization equilibrium, and radiative cross-sections, which through the radiative transport give form  to the thermal radiation emerging from the stellar surface.
The study of atoms in high magnetic field is, therefore, essential for interpretation of emergent spectra from atmospheres and radiating surfaces of compact stars \citep{kulebi:2009, potekhin:2014}. It also has considerable interest for solid state physics \citep{elliott:1960, orton:2004, bartnik:2010} and fundamental physics \citep{friedrich:1989}.

The effects of magnetic fields over the atomic structure are usually measured with the parameter $\beta=B/B_0$, where $B_0\approx 4.70103\times10^9$~G is set by equating the Bohr radius ($a_\text{B}=\hbar^2/m_\text{e}c^2$) to the characteristic magnetic length $a_\text{M}=\sqrt{2\hbar c/eB}$ (the mean value of the cyclotron radius in the lowest Landau state).  Thus, $\beta\ll1$ compresses the regime of weak magnetic fields, $\beta\approx 1$ that one where Coulomb and magnetic forces in the atom have comparable strengths, and $\beta\gg1$ the domain of the magnetic field on particle dynamics transverse to the field direction.

The properties of hydrogen atoms in an external magnetic field have been the subject of investigations over many decades \citep{garstang:1977, johnson:1983, lai:2001, thiru:2014}. It has already been recognized in studies of excitons (pairs of electron and hole) in semiconductors \citep{elliott:1960,hasegawa:1961} that, for very strong fields ($\beta\gg1$), the energy eigenfunctions can be approximated by the product of a Landau orbital (an energy eigenstate corresponding to a free-electron in the field $B$) and a function depending on the coordinate parallel to the field. This is the so-called adiabatic approximation introduced by \citet{schiff:1939}, which becomes exact in the limit $\beta\rightarrow\infty$, where the behaviour of the (non-relativistic) atomic states is reproduced by analytical results based on the one-dimensional hydrogen atom \citep{loudon:1959, haines:1969, loudon:2016}. 

Motivated by the discovery of pulsars ($\beta\ga1$) and magnetic white dwarfs ($\beta\la1$), energy evaluations were obtained with variational techniques using trial wavefunctions \citep{cohen:1970, smith:1972}, followed by more detailed numerical calculations at $\beta\gg1$ using the adiabatic approximation (e.g., \citet{canuto:1972}). Corrections to previous results were progressively obtained up to reach a comprehensive account of the first low-energy levels from appropriate wavefunction expansions in terms of spherical harmonics ($\beta\la1$) or Landau states ($\beta\ga1$) \citep{rosner:1984}. 
More recently, some authors (e.g., \citet{kravchenko:1996}) found very high accurate, nonrelativistic solutions for a few H states in constant fields of arbitrary strength. 
Presently, high precision values are known for a big number of energy states over the whole magnetic field strengths using a wavefunction expansion in terms of B-splines \citep{schi:2014}.

The above cited studies assumed an infinite nuclear mass and neglected the motion of the atoms. However, astrophysical fluid models require taking finite temperatures and hence the thermal motion of particles into account.
\citet{pavlov:1980} found simple energy scaling rules  to account the effects of finite proton mass for atoms at rest, but considerations of the particle movement effects are more complex. Motion perpendicular to a magnetic field breaks the axial symmetry that characterizes to an atom in rest and makes this problem fully three-dimensional.
\citet{gorkov:1968} showed that the motion of an atom across a magnetic field affects its internal energies. They found a conserved quantity, the so-called {\it pseudo-momentum} $\bm{K}$ introduced by \citet{johnson:1949} for single charges, with which the separation of the center-of-mass (CM) motion to the relative electron-proton motion is possible. At large enough values of the pseudo-momentum transverse to the magnetic field ($k_\perp$), or equivalently at a large electric field crossed to the magnetic one, the wave function of the relative motion is shifted from the Coulomb center to a magnetic well \citep{burkova:1976}. These states are called {\it decentered} states and are characterized by a large dipole moment as a result of the corresponding separation between the electron and proton. \citet{ipatova:1984} showed that the change in the dependence of the atom energy on the pseudo-momentum from centered (the wave function concentrated near the Coulomb well) to decentered states is rather abrupt and occurs when $k_\perp$ reaches certain critical value $\cm{K}_\text{c}$. Perturbative calculations of this dependence at $k_\perp\ll \cm{K}_\text{c}$ were implemented by \citet{vincke:1988} and \citet{pavlov:1993}. Non-perturbative results have been given in different works \citep{vincke:1992, lai:1995, potekhin:1998, lozovik:2004, potekhin:2014b}.

The ionization equilibrium of hydrogen in strong magnetic fields was first discussed by \citet{gnedin:1974}. \citet{khersonskii:1987,khersonskii:1987b} improved previous study by taking into account quantization of protons and finite nuclear mass in the atomic internal states. The influence of the pseudo-momentum on the atomic partition function was considered by \citet{ventura:1992} but without providing quantitative results. \citet{pavlov:1993} included changes of the internal atomic structure caused by the thermal motion of the atoms across the magnetic field, using perturbative evaluations at $\beta\ga 1$. Approximate evaluations of the hydrogen ionization equilibrium including pseudo-momentum effects were considered at \citet{lai:1995} for superstrong fields ($\beta\gg1$), and improved results have been then given by Potekhin and coworkers (e.g., \citet{potekhin:1999,potekhin:2014b}) also focused on the very intense magnetic fields of neutron star atmospheres ($B\ga10^{10}$~G).

The present paper is aimed to evaluate the ionization equilibrium of magnetized hydrogen atoms in the region of intermediate values of magnetic field (mainly $B\approx 10^5$~--~$10^{10}$~G), which have remained unexplored so far. For this purpose, we write fitting formulae for the energies of bound states in the transition between field-free and high-field regimes using accurate numerical data for atoms at rest, combined with analytical results of CM effects due to thermal particle motions. 
On the other side, we develop mathematical relations to express the correspondence between states in the field-free and strong-field regimes. Our study is focused to conditions found in the atmospheres of magnetic white dwarf stars with megagauss fields ($B\approx 10^6$~--~$10^{9}$~G), for which zero-field occupation numbers of atoms and ions  are currently used \citep{eucher:2002, aznar:2004, kulebi:2009}.

The paper is organized as follows. Section \ref{s.atoms} shows the quantum problem of a hydrogen atom in an uniform magnetic field. Section \ref{s.states} is devoted to establish simple rules for the relationships between atomic states in the weak and strong field limits. Section \ref{s.energies} reviews the binding energies of atoms in a magnetic field on different regimes including the treatement of finite nuclear mass and moving particle effects. Section \ref{s.stat} summarizes the chemical potentials, partition function and ionization equilibrium equations. Results and their analysis are shown in Section \ref{s.results}. Concluding remarks are given in Section \ref{s.concl}.

%%%%%%%%%%%%%%%%%%%%%%%%%%%%%%%%%%%%%%%%%%%%%%%%%%%%%%%%%%%%%%%%%
\section{Hydrogen atoms in a magnetic field}\label{s.atoms}

The Hamiltonian of the hydrogen atom in a magnetic field omitting relativistic effects is
\beq
H= \frac{\pi_\text{p}^2}{2m_\text{p}} +\frac{\pi_\text{e}^2}{2m_\text{e}} +V(r)
\eeq
with 
\beq \label{e.pi}
\bm{\pi}_\text{i} = \bm{p}_\text{i}-\frac{q_\text{i}}{c} \bm{A},
\eeq
where $q_\text{i}$, $m_\text{i}$, $\bm{r}_\text{i}$, $\bm{p}_\text{i}$ and $\bm{\pi}_\text{i}$ are respectively the charge ($q_\text{e}=-e$, $q_\text{p}=e$), 
mass, position, canonical momentum and kinetic momentum of the electron ($\text{i}=\text{e}$) or proton ($\text{i}=\text{p}$); $\bm{A}$ is the potential vector of the field and $V(r)=-e^2/r$ the Coulomb potential with $r$ the magnitude  of $\bm{r}=\bm{r}_\text{e}-\bm{r}_\text{p}$.

For homogeneous magnetic fields $\bm{B}$, the motion integral of the atom related with the translational invariance of the Hamiltonian is given by the  pseudo-momentum operator $\bm{K}$ \citep{gorkov:1968, avron:1978, herold:1981, johnson:1983}
\beq \label{e.Kdef}
\bm{K}= \bm{\pi}_\text{e} +\bm{\pi}_\text{p} -\frac{\bm{b}\times\bm{r}}{2},
\eeq
with $\bm{b}=e\bm{B}/c$ and the gauge
\beq \label{e.A}
\bm{A}(\bm{r}) = \frac{\bm{B}\times\bm{r}}2.
\eeq
$\bm{K}$ is useful to separate the relative motion of the electron and proton from the mass center motion \citep{gorkov:1968}. The eigenenergy equation of the atom in relative coordinates can be written in the form \citep{herold:1981}
\beq \label{e.Hrel}
\left[ \frac{\pi^2}{2\mu} +\frac{(\bm{k}+\bm{b}\times\bm{r})^2}{2M}
  +V(r) \right]\phi(\bm{r}) =E\phi(\bm{r}),
\eeq
with  $\phi(\bm{r})$ the eigenfunction of Hamiltonian and 
\beq
\bm{\pi}=\bm{p}+\frac{\gamma(\bm{b}\times\bm{r})}{2},
\eeq
$\bm{k}$ being an eigenvalue of the pseudo-momentum operator $\bm{K}$, $\bm{p}$ the one-particle canonical momentum ($\bm{p}=-i\hbar\nabla_\mb{r}$ in the relative coordinate space), $M$ the total mass, $\mu$ the reduced mass, and $\gamma$ a relative mass difference, 
\beq
M= m_\text{p} + m_\text{e},\hskip.4in
\mu=\frac{m_\text{e} m_\text{p}}{M},\hskip.4in
\gamma= \frac{m_\text{p}-m_\text{e}}{M}.
\eeq

Eq. (\ref{e.Hrel}) has been largely studied for atoms in rest ($\bm{k}=0$), where the Hamiltonian of the relative motion of electron and proton takes the form
\beq \label{e.Harest}
H_\text{rel} = \frac{p^2}{2\mu} +\frac{\gamma\mb{b}.\mb{L}}{2\mu}
  +\frac{(\mb{b}\times\mb{r})^2}{8\mu} +V(r),
\eeq
$\mb{L}=\mb{r}\times\mb{p}$ being the relative angular momentum operator.

Hereafter, we adopt the $z$ axis of the cartesian and cylindrical coordinates oriented in the $\bm{B}$ direction.

%%%%%%%%%%%%%%%%%%%%%%%%%%%%%%%%%%%%%%%%%%%%%%%%%%%%%%%%%%%%%%%%%
\subsection{States in the zero- and strong-field limits}\label{s.limits}

A brief review of atomic states in the limits $\beta\rightarrow 0,\infty$ is required to specify the correspondence between both of them. When the magnetic field is switch off, Eq. (\ref{e.Harest}) reduces to the Hamiltonian of the usual Coulomb problem. Bound states are typically represented by eigenstates $\phi_{n,l,m}$ common to the Hamiltonian, $L^2$ and $L_z$ operators, which have eigenvalues $-E_\text{H}/n^2$, $\hbar^2l(l+1)$, and $\hbar m$, respectively, depending on the principal ($n=1,2,\dots$), orbital ($l=0,1,\dots,n-1$) and magnetic ($m=-l,\dots,l-1,l$) quantum numbers. $E_\text{H}=13.605693$~eV is the ionization energy of the field-free atom. 

At very intense field values ($\beta\gg1$), Coulomb interaction has a negligible effect on the (electron and proton) relative motion transverse to the magnetic field. Eigenfunctions of $H_\text{rel}$ can be then factorized in the product \citep{schiff:1939}.
\beq
\Phi = \Phi^\perp(r_\perp,\varphi)\, \Phi^\parallel (z),
\eeq
with separated dependence on cylindrical coordinates ($r_\perp,\varphi,z$), while the eigenenergies are expressed as a sum
\beq \label{e.EpEp}
E=E^\perp+E^\parallel.
\eeq
Due to the presence of the magnetic field, $L^2$ is no longer a conserved quantity but the Hamiltonian still commutes with $L_z$, so that $m$ remains as a good quantum number. The transversal part of the eigenfunction is given by the usual Landau function $\Phi^\perp_{N,m}$ labeled by $m$ and the Landau number $N$, while the longitudinal component $\Phi^\parallel_\nu$ of the wavefunction is determined by an effective potential parallel to the field, with $\nu$ the {\it longitudinal} quantum number which takes non-negative integer values for negative values of $E^\parallel$ (in such case, $\nu$ gives the number of nodes of $\Phi^\parallel_\nu$) and it is continuous otherwise.

With the zero-point and spin terms subtracted, the transverse contribution to the eigenenergy in Eq. (\ref{e.EpEp}) can be written as
\beq
E^\perp =\hbar\omega_\text{e} N + \hbar\omega_\text{p}(N-m),
\eeq
with
\beq\label{e.omega}
\omega_\text{i}=\frac{eB}{m_\text{i} c},
\eeq
and
\beq\label{e.Nnr}
N=n_\text{r} +\frac{|m|+m}2 ,
\eeq
$n_\text{r}$ ($=0,1,2,\dots$) being the {\it radial} quantum number which enumerates the nodes of $\Phi^\perp_{N,m}$ along the coordinate $r_\perp$. Because $n_\text{r}\ge0$ and $m\le N$ ($=0,1,2,\dots$), $E^\perp$ is positive or zero.

Binding values of the longitudinal energy ($E^\parallel<0$) exist for $m\le0$ and were calculated with the one-dimensional hydrogen atom approach \citep{loudon:1959, haines:1969}. Its solutions are composed by tightly bound states ($\nu=0$) and hydrogen-like states ($\nu=1,2,3\dots$). The longitudinal energies of tightly states can be approximated by \citep{ruderman:1974}
\beq \label{E.adv0}
E^\parallel_{\nu=0}=-0.32E_\text{H}\ln^2\left(\frac{2\beta}{2|m|+1}\right),
\eeq
and those of hydrogen-like states converge to a Rydberg series
\beq \label{E.adv1}
E^\parallel_{\nu>0}=-E_\text{H} \left[ \text{Int}
 \left(\frac{\nu+1}2\right) + \delta_{\nu m}\right]^{-2},
\eeq
with $\text{Int}(x)$ the integer part of $x$, and $\delta_{\nu m}$ a quantum defect parameter which takes negative values and vanishes in $\beta\rightarrow\infty$ \citep{friedrich:1989}, e.g., $\delta_{\nu=1,m}=-4\left(\frac{2|m|+1}{2\beta}\right)^{1/2}$.

Solutions (\ref{E.adv0}) and (\ref{E.adv1}) correspond to a fixed Coulomb potential (i.e., infinitely massive proton). 
If finite nuclear-mass effects are ignored, the only bound states below the first Landau level ($N=0$) are those with $m\le0$ ($n_\text{r}=0$). States over the first Landau level, which determines the edge of the continuum energy, are metastable ($m<N$, $n_\text{r}>0$) or truly bound ($m=N$, $n_\text{r}=0$) depending on the existence or not of lower states with the same magnetic quantum number \citep{simola:1978}.
When the finite proton mass is taking into account only states with $N=0$ and $m=0$ remain below the continuum edge at $\beta\rightarrow\infty$ (see Eq. (\ref{e.finite_mass})).

%%%%%%%%%%%%%%%%%%%%%%%%%%%%%%%%%%%%%%%%%%%%%%%%%%%%%%%%%%%%%%%%%
\section{Bound state correspondence}\label{s.states}

At very weak magnetic fields ($\beta\ll1$) the atomic states are well described by the Coulomb quantum numbers $\{n,l,m\}$. The Zeeman quadratic effect removes the $l$ degeneracy of low-lying states in $\beta\la10^{-3}$, while $n$ remains a good quantum number. Inter-$n$ mixing starts to appear at highly excited states for very low fields and reaches the lowest states at $\beta\approx1$. In the $n$ mixing regime, the energy level pattern is characterized by many close anti-crossings \citep{ruder:1994, schi:2014}. This complex structure disappears for strong magnetic fields ($\beta\gg 1$) where a new ordered structure arises formed by Rydberg-like levels plus tightly bound states. There, the set of quantum numbers $\{N,\nu,m\}$ 
becomes appropriate to describe the atomic states. Two quantities are conserved on the whole range of magnetic fields, the $z$-parity ($\pi_z$) of the energy eigenfunctions and the $z$-component of the orbital angular momentum (quantum number $m$). Consequently, $\pi_z$ and $m$ remains as good quantum numbers in arbitrary field intensity. As it is well-known, the longitudinal parity is $\pi_z=(-1)^{l-m}$ for free-field states and $\pi_z=(-1)^\nu$ for bound states in the strong field regime.

The correspondence between energy states at low and high fields was clarified by \citet{simola:1978} using the non-crossing rule of states. With this rule, which applies to states with exactly the same symmetries on the Hamiltonian operator, bound states in both limits ($\beta\rightarrow 0,\infty$) corresponding to the same $\pi_z$ and $m$ are connected in the order of growing energy. Nevertheless, there are ($N,\nu,m$) states which remain at $\beta\rightarrow0$ as a linear combination of two or more $(n,l,m)$ states with the same $n$ and $m$ but different $l$, one of these states usually being the dominant one. 
In the practice, for degenerate ($n,m,\pi_z$) multiplets one can adopt the state with highest $l$ which has the lowest energy at $\beta\ll1$ \citep{ruder:1994}. Adopting this convention, it is possible to establish a complete one-to-one correspondence between the field-free states and strong-field states.

%=============================================
\begin{table*}
\caption{Longitudinal quantum number $\nu$ of high-field states connected to zero-field states ($n,l,m$), for levels $1\le n \le 5$ and $n=20$. \label{T.1}}
\setlength{\tabcolsep}{3pt}
\begin{small}
\begin{tabular}{rrrr rrrr rrrr rr rr rrrrr rrrrr rrrrr r}  
\hline 
 $n$ & $l$ & & $|m|=0$ & 1 & 2 & 3 & 4 & $\hskip.1in$ &
 $n$ & $l$ & & $|m|=0$ & 1 & 2 & 3 & 4 & 5 & 6 & 7 & 8 & 9 
& 10 & 11 & 12 & 13 & 14 & 15 & 16 & 17 & 18 & 19 \\ \hline
     &    &&    &&&&& &  20 &  0& & 218& & & &
& &&&&&&&&&&&&& \\
   1 &  0 &&  0 &&&&& &  20 &  1& & 199& 198& & &
& &&&&&&&&&&&&& \\
    &&&& &&&& &  20 &  2& & 216& 179& 178& & 
& &&&&&&&&&&&&& \\
   2 &  0 &&  2 &&&&& &  20 &  3& & 197& 196& 161& 160&
& &&&&&&&&&&&&& \\
   2 &  1 &&  1 &  0 &&&& & 20  &  4& & 214& 177& 176& 143& 142&
& &&&&&&&&&&&&& \\
    &&&& &&&& &  20 &  5& & 195 & 194& 159& 158& 127&
126& &&&&&&&&&&&&& \\
   3 &  0 &&  6 &&&&& &  20 &  6& & 212& 175& 174& 141& 140&
111& 110& &&&&&&&&&&&& \\
   3 &  1 &&  3 &  2 &&&& &  20 &  7& & 193& 192& 157& 156& 125&
124&  97& 96& &&&&&&&&&&& \\
   3 &  2 &&  4 &  1  & 0 &&& &  20 &  8& & 210& 173& 172& 139& 138&
109& 108& 83& 82& &&&&&&&&&& \\
    &&&& &&&& &  20 &  9& & 191& 190& 155& 154& 123&
122&  95& 94& 71& 70& &&&&&&&&& \\
   4 &  0 && 10 &&&&& &  20 & 10& & 208& 171& 170& 137& 136&
107& 106& 81& 80& 59& 58& &&&&&&&& \\
   4 &  1 &&  7 &  6 &&&& &  20 & 11& & 189& 188& 153& 152& 121& 
120&  93& 92& 69& 68& 49& 48& &&&&&&& \\
   4 &  2 &&  8 &  3 &  2 &&& &  20 & 12& & 206& 169& 168& 135& 134&
105& 104& 79& 78& 57& 56& 39& 38& &&&&&& \\
   4 &  3 &&  5 &  4 &  1 &  0 && &  20 & 13& & 187& 186& 151& 150& 119& 
118&  91& 90& 67& 66& 47& 46& 31& 30& &&&&& \\
    &&&& &&&& &  20 & 14& & 204& 167& 166& 133& 132&
103& 102& 77& 76& 55& 54& 37& 36& 23& 22& &&&& \\
   5 &  0 && 16 &&&&&   &  20 & 15& & 185& 184& 149& 148& 117&
116&  89& 88& 65& 64& 45& 44& 29& 28& 17& 16& &&& \\
   5 &  1 && 11 & 10 &&&& &  20 & 16& & 202& 165& 164& 131& 130&
101& 100& 75& 74& 53& 52& 35& 34& 21& 20& 11& 10& && \\
   5 &  2 && 14 &  7  & 6 &&& &   20 & 17& & 183& 182& 147& 146& 115& 
114&  87& 86& 63& 62& 43& 42& 27& 26& 15& 14&  7& 6& & \\
   5 &  3 &&  9 &  8  & 3 &  2 && &  20 & 18& & 200& 163& 162& 129& 128&
 99&  98& 73& 72& 51& 50& 33& 32& 19& 18&  9&  8& 3& 2& \\
   5 &  4 && 12 &  5  & 4 &  1 &  0 & &  20 & 19& & 181& 180& 145& 144& 113&
112&  85& 84& 61& 60& 41& 40& 25& 24& 13& 12&  5& 4& 1& 0 \\
\hline
\end{tabular}
\end{small}
%\tablefoot{...}
\end{table*}
%=============================================
Following the previous convention, we found simple relatioships connecting the sets of quantum numbers of {\it bound} states in field-free and high intensity field regimes. In particular, one can express $\nu$ in terms of Coulomb quantum numbers in a straightforward way. 
For states with positive parity $\pi_z=+1$, i.e. ($l+m$) and $\nu$ even,
\beq \label{e.nu_even}
\nu =\left\{ \begin{array}{ll}\displaystyle
\frac{1}{2}\left[(n-|m|+1)^2-4\right]-l+|m|,   &\quad \text{odd}~(n+m), \\ 
[2ex]\displaystyle
\frac{1}{2}\left[(n-|m|+1)^2-5\right]-l+|m|,   &\quad \text{even}~(n+m).\\
\end{array} \right.
\eeq
For states with negative parity $\pi_z=-1$, i.e. ($l+m$) and $\nu$ odd,
\beq \label{e.nu_odd}
\nu = \left\{ \begin{array}{ll}\displaystyle
\frac{1}{2}\left[(n-|m|)^2-1\right]-l+|m|,     &\quad \text{odd}~(n+m), \\ 
[2ex]\displaystyle \frac{1}{2}(n-|m|)^2-l+|m|, &\quad \text{even}~(n+m).\\
\end{array} \right.
\eeq
The correlation between bound states is closed with the relation
\beq \label{e.N_m}
N = \left\{ \begin{array}{ll}\displaystyle
0, &\quad (m\le0), \\ 
m, &\quad (m>0).\\
\end{array} \right.
\eeq
Eqs. (\ref{e.nu_even})-(\ref{e.N_m}) give a complete correspondence between ($n,l,m$) and ($N,\nu,m$) bound states.
This quantitative scheme generalizes examples shown in \citet{ruder:1994}. Table \ref{T.1} lists the sublevel correspondence to high-field for a few low $n$ and a highly excited state ($n=20$) in the zero-field limit. 

When the field is switch off, the energy of a bound state ($N,\nu,m$) converges to a Bohr level. In agreement with relationships (\ref{e.nu_even}) and (\ref{e.nu_odd}), the main quantum number $n$ is given by
\beq \label{e.n_v}
n= \left\{ \begin{array}{ll}\displaystyle
\text{Int}\left[1+\frac{2\nu}{1+\sqrt{2\nu+1}}\right]+|m|, & \quad (\text{even}~\nu),\\
[3ex]\displaystyle 
\text{Int}\left[2+\frac{2\nu-2}{1+\sqrt{2\nu-1}}\right]+|m|, & \quad (\text{odd}~\nu).\\
\end{array} \right. 
\eeq
This last state correlation takes into account the superposition of degenerate angular momentum states ($l$) inside of a ($n,m,\pi_z$) multiplet and, therefore, it is independent of the adopted convention of one-to-one correspondence. 

The state correspondence ansatz given here provides a suitable scheme for evaluating the atomic partition function over the whole range of magnetic fields. Specifically, this let us freely move through both representations (\{$n,l,m$\} and \{$N,\nu,m$\}).

%%%%%%%%%%%%%%%%%%%%%%%%%%%%%%%%%%%%%%%%%%%%%%%%%%%%%%%%%%%%%%%%%
\section{Bound state energies} \label{s.energies}

The second term in the square brackets of Eq. (\ref{e.Hrel}) couples translational and internal energies of the H atom throught the tranverse component ($\bm{k}_\perp$) of the pseudo-momentum
\beq \label{e.k}
\bm{k} = \bm{k}_\perp + \bm{k}_z.
\eeq
It is worth noting that $\bm{k}_z=\bm{p}_z$, which follows from Eqs. (\ref{e.pi})-(\ref{e.A}).
Differents approximations are used for low and high $\bm{k}_\perp$ values.

%%%%%%%%%%%%%%%%%%%%%%%%%%%%%%%%%%%%%%%%%%%%%%%%%%%%%%%%%%%%%%%%%
\subsection{Centered states}\label{s:at_cen}

For small $\bm{k}_\perp$, the coupled term in the Hamiltonian can be treated as a perturbation and the eigenenergies at second order are written as \citep{vincke:1988, pavlov:1993}
\beq \label{E.0}
E=\cm{E} +\frac{k_\perp^2}{2M_\perp} +\frac{k_z^2}{2M},
\eeq
where $\cm{E}$ is the energy of the rest atom, i.e., a solution of Eq. (\ref{e.Harest}), and $M_\perp$ an effective mass given by 
\beq \label{e.Mtrav}
M_\perp = \frac{M}{1-\alpha},
\eeq
with
\beq \label{e.delta}
\alpha \approx \hbar\omega_\text{H}
 \left(  \frac{1-m}{\cm{E}_{\nu,m-1}-\cm{E}_{\nu,m}+\hbar\omega_\text{p}}
        - \frac{m}{\cm{E}_{\nu,m+1}-\cm{E}_{\nu,m}-\hbar\omega_\text{p}}
 \right).
\eeq
Velocity effects on the energy spectrum become of significant importance for large enough fields, particularly in the regime $\beta\gg1$ where Eq. (\ref{e.delta}) has been derived using a basis of states $\{N,\nu,m\}$. 
At very low field intensities ($\beta\la1$), a perturbative method based on states $\{n,l,m\}$ would be more appropriate, however, current expression for $\alpha$ is well behaved in this regime (where its effects are significantly reduced) so that we use that approach on the whole range of $\beta$.
The application of the perturbation method demands energy corrections lower than the spacing of adjacent unperturbed levels (e.g., $\Delta\cm{E}=|\cm{E}_{\nu,m}-\cm{E}_{\nu,m+1}|$ in large $\beta$). Consequently, this gives an upper limit to the magnitude of the  transverse pseudo-momentum \citep{pavlov:1993}
\beq
k_\perp \ll \sqrt{ \frac{2M \Delta\cm{E}}{\alpha}}.
\eeq
States for which the approximation given by Eq. (\ref{E.0}) is valid are often called {\it centered states} (even when them could be weakly decentered states), because the electron wavefunction remains on average centered around the Coulomb well. In present work, anisotropy mass ($M_\perp$) has been calculated with Eqs. (\ref{e.Mtrav}) and (\ref{e.delta}) using energy data of \citet{schi:2014}. Some fits are given in Section \ref{s.mass}.

Solutions to Eq. (\ref{e.Harest}) are usually found in the approximation of infinite nuclear mass ($\mu\rightarrow m_\text{e}$, $\gamma\rightarrow 1$), which are here denoted by $\cm{E}_\infty$. The eigenvalues $\cm{E}$ for finite mass may be then evaluated from the scaling relations \citep{pavlov:1980}
\beq \label{e.finite_mass}
\cm{E}=\frac{\mu}{m_\text{e}}\cm{E}_\infty\left[\left(\frac{m_\text{e}}{\mu}\right)^2\beta\right]-\left(\frac{m_\text{e}}{\mu}\right)^2 \hbar \omega_\text{p}(m+m_s),
\eeq
with $m_s=\pm1/2$ the spin quantum number of the electron. Notice that Eq. (\ref{e.finite_mass}) includes a scaling relation for the field intensity where $\cm{E}_\infty$ is evaluated. For the present case (electron and proton pairs) $\mu\approx m_\text{e}$ and only the second term on the r.h.s. of Eq. (\ref{e.finite_mass}) is truly significant in occupation number calculations.

%=============================================
\begin{figure}
\includegraphics[width=.48\textwidth]{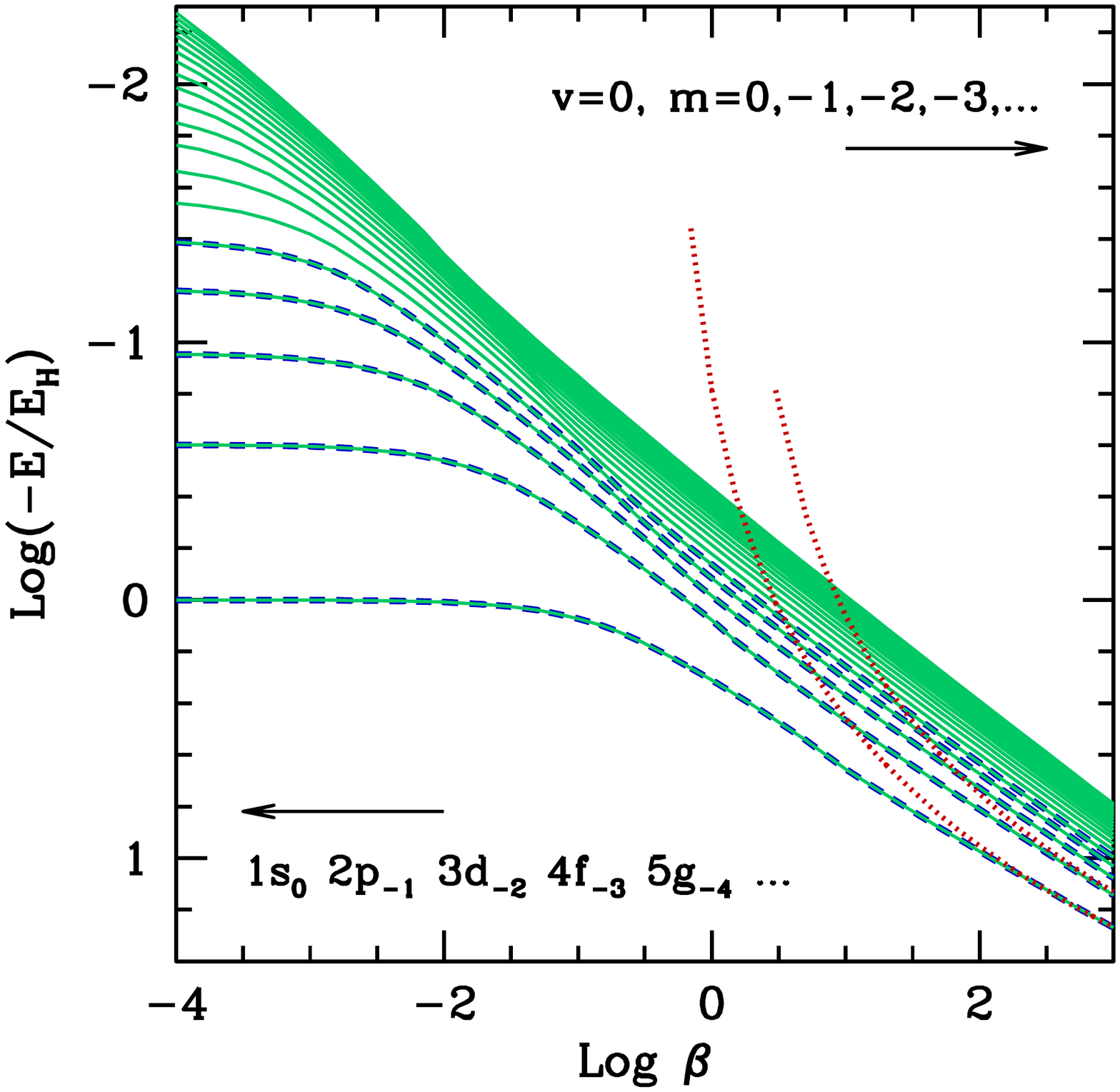}
\caption{Energies of tighly bound states ($\nu=0$) of atoms in rest and infinity nuclear mass, as a function of the magnetic field. Dashed lines: numerical results from \citet{schi:2014} for the lowest five states ($0\ge m\ge-5$) with spin-down. Solid lines: our fits and extrapolations up to $m=-19$. Dotted lines: the asymptotic approximations given by Eq. (\ref{E.adv0}) for $m=0,-1$.\label{f:v0}}
\end{figure}
%=============================================
%=============================================
\begin{figure}
\includegraphics[width=.48\textwidth]{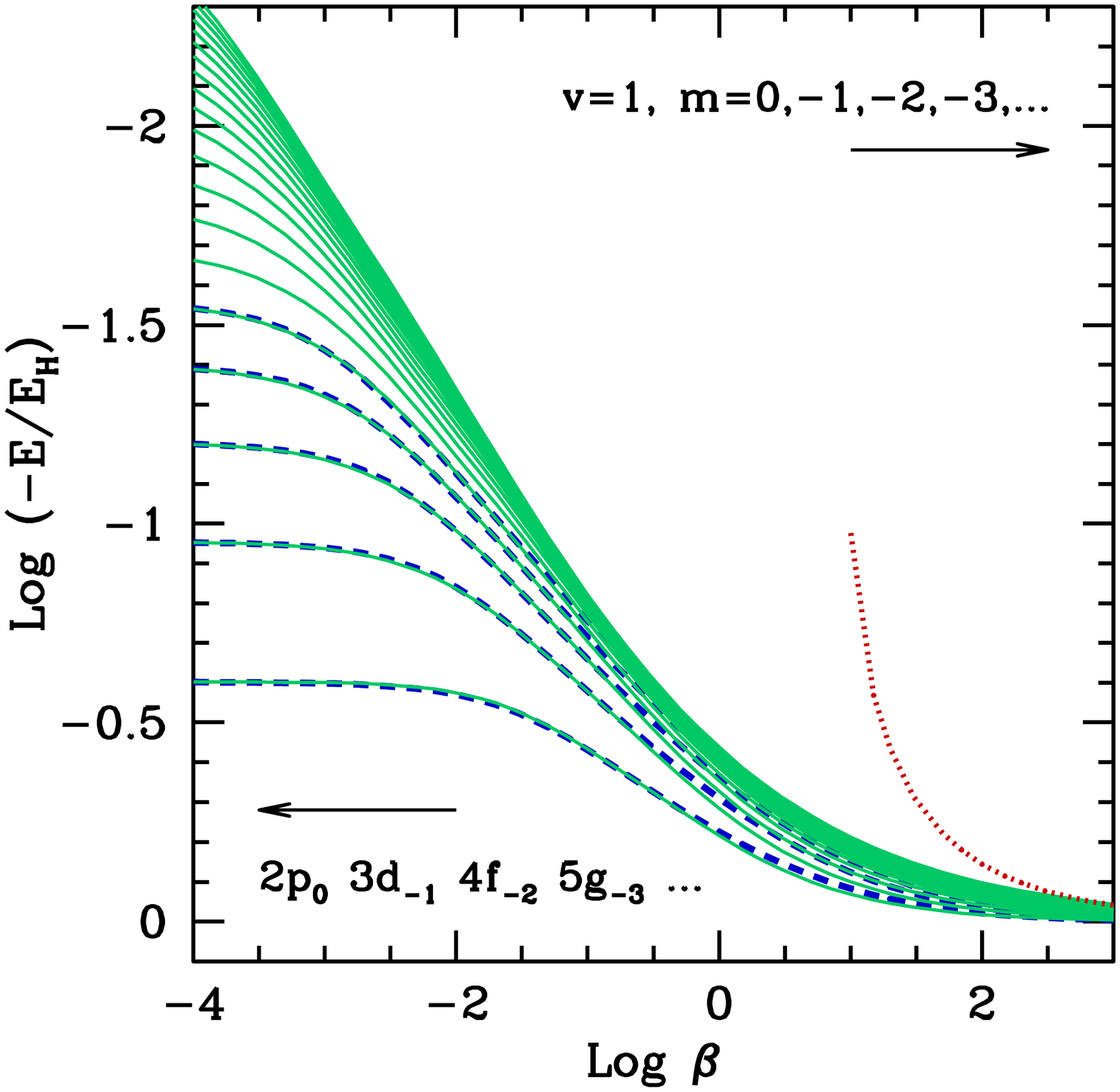}
\caption{Idem Fig. \ref{f:v0} but for hydrogen-like states at $\nu=1$. 
Dotted line corresponds to predictions of Eq. (\ref{E.adv1}) for $m=0$.
\label{f:v1}}
\end{figure}
%=============================================
In present work, we adopt the energies $\cm{E}_\infty$ calculated by \citet{schi:2014} which comprise spin-down ($m_s=-1/2$) levels emerging from zero-field states with principal quantum numbers $n\le15$ and magnetic quantum numbers $-4\le m\le 0$. Energy values of states with $m_s=+1/2$ and $m>0$ are respectively obtained by adding $4\beta$ and $4m\beta$ to values $\cm{E}_\infty(m<0,m_s=-1/2)$ measured in Rydbergs. 
Results from \citet{schi:2014} let us find scaling relations for the energy dependence on $m$ and $\nu$ and approximate energy curves for arbitrary bound states (see Appendix \ref{a:1}). For instance, Figs. \ref{f:v0} and \ref{f:v1} show $\cm{E}_\infty$ curves at $\nu=0$ and $\nu=1$ coming from the first twenty Bohr levels. These fits are valid for any $B$ with relative errors lower than one percent and converge to the right limits at zero-field and very strong fields. 
Typical well-known energy expressions for strong fields (Eqs. (\ref{E.adv0})-(\ref{E.adv1})) show strong departures at $\beta\la 50$ for the ground state and at much larger $\beta$ for excited states. In particular, they can not be used in the regime of magnetic white dwarfs ($-6.6\la\log\beta\la-0.6$). 

The accuracy of our energy fits varies for different levels and different field strengths. The mean relative error ($\sigma_E$) for the ground state does not exceed 0.4\% on the range $-4\le\log\beta\le3$ and falls below 0.1\% in the MWD region. Other tightly bound states ($\nu=0$, $-4\le m\le-1$) have relative errors of few tenths of percent in the MWD domain. General expression for $\nu\ge4$ provides energies with $\sigma_E$ typically between two and four percent, except peaks ($\approx7$\%) in some few hydrogen-like states (e.g., $\nu=4, 8, 12$).
%=============================================
\begin{figure}
\includegraphics[width=.48\textwidth]{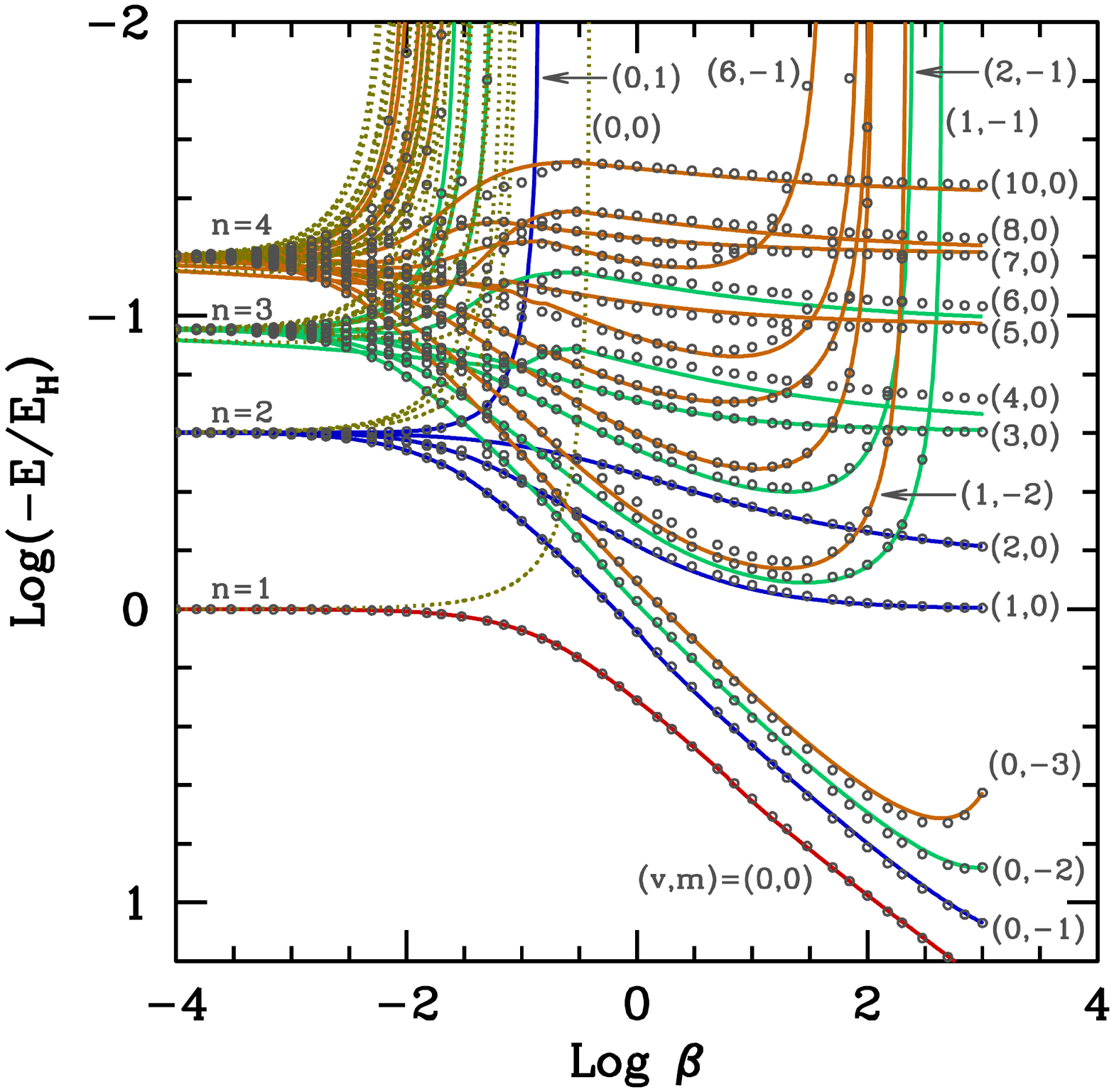}
\caption{Energies of magnetized atoms at rest coming from Bohr levels $n=1$, $2$, $3$ and $4$, as a function of the field intensity. Finite nuclear mass effects are included. Lines represent fits used in this work for spin-down (solid) and spin-up (dotted) states. Symbols correspond to spin-down values calculated by \citet{schi:2014}. Some ($\nu,m$) states are indicated on the plot.
\label{f:EHk0}}
\end{figure}
%=============================================
A partial energy spectrum of the atom at rest taking into account finite nuclear mass is appreciated in Fig. \ref{f:EHk0}. This shows the dependence with the field intensity of the states arising from the first four Bohr levels, and a comparison of our results from analytical expressions (lines) with \citet{schi:2014} evaluations for spin-down states (symbols). The ionization energy of the atom, represented by the negative value of the solid curve labeled $(\nu,m)=(0,0)$, increases monotonically with the field strength. As a consequence of the positive energy coupling with the field, spin-up states and those with positive magnetic quantum number increase their energies and move toward the continuum at relatively low field intensities, $0.01<\beta<1$. A second state migration toward the continuum occurs at high fields ($\beta>10$, for the states showed in Fig. \ref{f:EHk0}), due to finite nuclear-mass corrections affecting states with $m<0$ (second term of the r.h.s. of Eq. (\ref{e.finite_mass})).
Only states with $m=0$ remain bound for any field strength.

%%%%%%%%%%%%%%%%%%%%%%%%%%%%%%%%%%%%%%%%%%%%%%%%%%%%%%%%%%%%%%%%%
\subsection{Decentered states}\label{s:at_decen}

At very large pseudo-momentum of the atom motion across the field, the electron probability density is markedly shifted apart from the Coulomb center and approaches to the so-called relative guiding center (separation between the electron and proton guiding centers)
\beq
\mb{r}_\text{c}=\frac{c}{eB^2}\mb{B}\times \mb{k},
\eeq
with magnitude
\beq \label{e.rc}
r_\text{c} = \frac{k_\perp a_\text{B}^2}{2\hbar \beta}
\eeq
($r_\text{c}=k_\perp/2\beta$ in atomic units).
In this case, the atomic state becomes {\it decentered} and the dependence of the energy levels on $k_\perp$ can not be interpreted in terms of a mass anisotropy.

The transition between weakly to strongly decentered states have been studied in the regime of strong magnetic fields $\beta\ga1$ \citep{vincke:1992, potekhin:1994}, where some useful analytical expressions have been given \citep{potekhin:1998, potekhin:2014b}. These expressions are adopted in the present work with minor changes to be extrapolated to low field intensity (for which we demand the condition $E\rightarrow\cm{E}$ at $k\rightarrow0$).
With energies measured in Rydbergs and $r_c$ in Bohrs, the energy of decentered states is approximated by
\beq \label{e.Ev0}
E=\frac{4m_e}{m_p}m\beta 
 +\frac{2}{\chi_0-\sqrt{r_c^2 +(2\nu+1)r_c^{3/2}+\chi_1}}
 +\frac{k_z^2}{2M},
\eeq
with
\beq \label{e.chi0}
\chi_0= \left\{ \begin{array}{ll}\displaystyle
0, & \quad (\nu=0),\\
[2ex]\displaystyle 
\frac{2}{\cm{E}_\infty}, & \quad (\text{otherwise}),\\
\end{array} \right. 
\eeq
and
\beq \label{e.chi1}
\chi_1= \left\{ \begin{array}{ll}\displaystyle
\frac{r_c}{5-3m}+\frac{4}{\cm{E}_\infty^2}, & \quad (\nu=0),\\
[3ex]\displaystyle 
\left[\nu^2+2^{0.5\nu \log(1+\beta/150)}\right]r_c, & \quad (\text{even}~\nu>0),\\
[2ex]\displaystyle 
\left(\nu^2-1\right)r_c, & \quad (\text{odd}~\nu).\\
\end{array} \right. 
\eeq
Eqs. (\ref{e.Ev0})-(\ref{e.chi1}) contain the expected asymptotic value of the energy for large transverse pseudo-momentum,
\beq \label{e.Ek_asymb}
E= \frac{4m_e}{m_p}m\beta  -\frac{4\beta}{k_\perp} + \frac{k_z^2}{2M},
\quad (k_\perp\rightarrow\infty),
\eeq
where the first term on the r.h.s. is a CM correction to the total energy, and the second term represents the Coulomb energy ($-e^2/r_\text{c}$) with the electron located on the magnetic well (i.e., to distance $r_\text{c}$ from the proton). On the other side, Eq. (\ref{e.Ev0}) converges to the expected result corresponding to low transverse motions 
\beq \label{e.Edecek0}
E=\frac{4m_e}{m_p}m\beta +\cm{E}_\infty + \frac{k_z^2}{2M} 
 \approx  \cm{E} + \frac{k_z^2}{2M},
\quad (k_\perp\rightarrow 0).
\eeq
In practice, following \citet{potekhin:2014b}, the transition region $k_\perp\approx \cm{K}_c$ between centered and decentered states is identified by the intersection of curves given by Eqs. (\ref{E.0}) and (\ref{e.Ev0}).
%=============================================
\begin{figure}
\includegraphics[width=.48\textwidth]{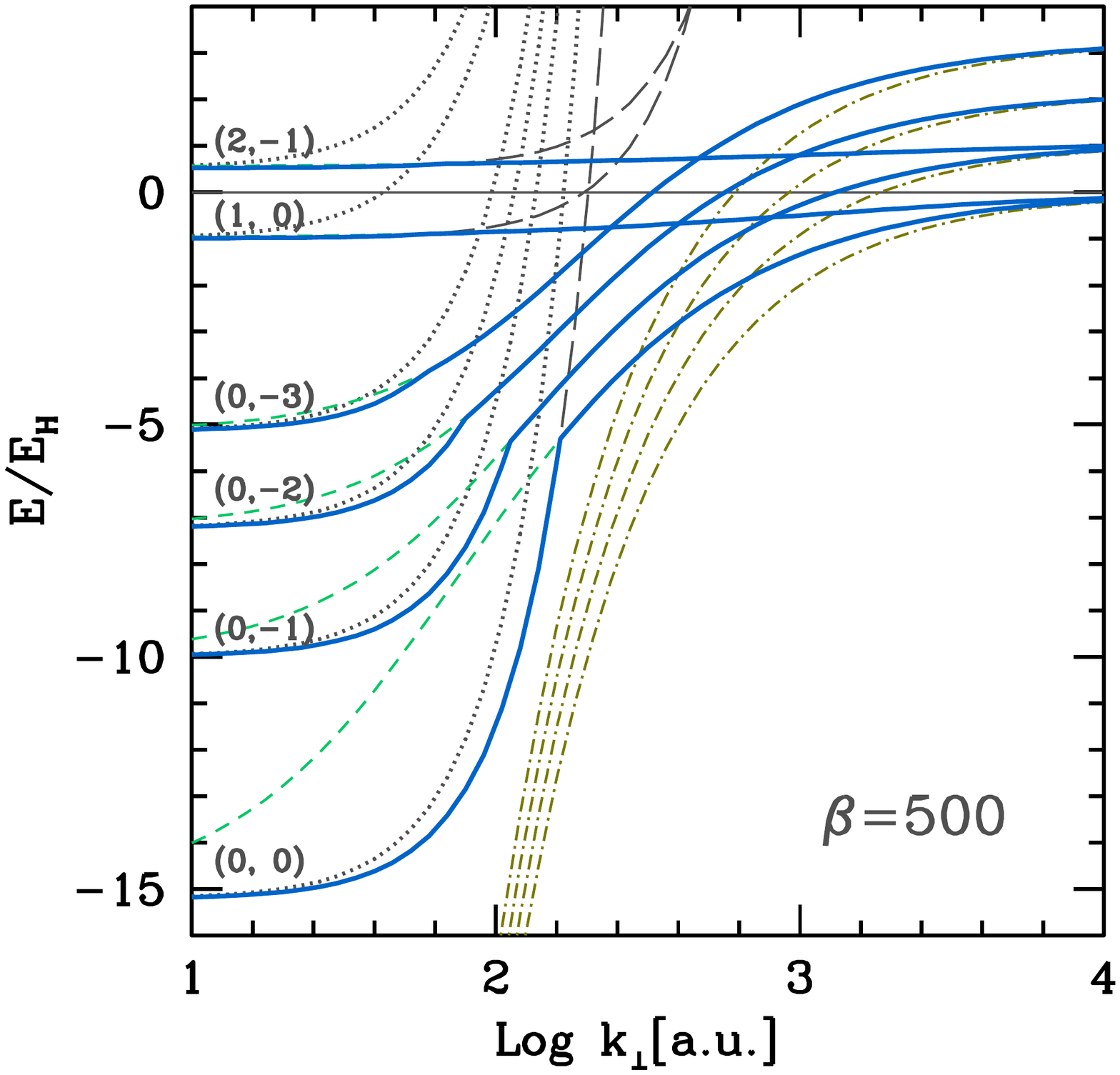}
\caption{Solid lines represent energies (without kinetic contribution from motion along the field) of magnetized atoms for a few $(\nu,m)$ states as a function of the transverse pseudo-momentum, at $\beta=500$ ($B=2.35\times 10^{12}$~G). 
Results without finite-velocity effects are indicated by dotted lines. Approximations based on Eqs. (\ref{E.0}), (\ref{e.Ev0}) and (\ref{e.Ek_asymb}) are displayed by short-dashed, long-dashed [only for $(0,0)$, $(1,0)$ and $(2,-1)$ states] and dot-dashed lines, respectively.
\label{f:EHk2}}
\end{figure}
%=============================================
%=============================================
\begin{figure}
\includegraphics[width=.48\textwidth]{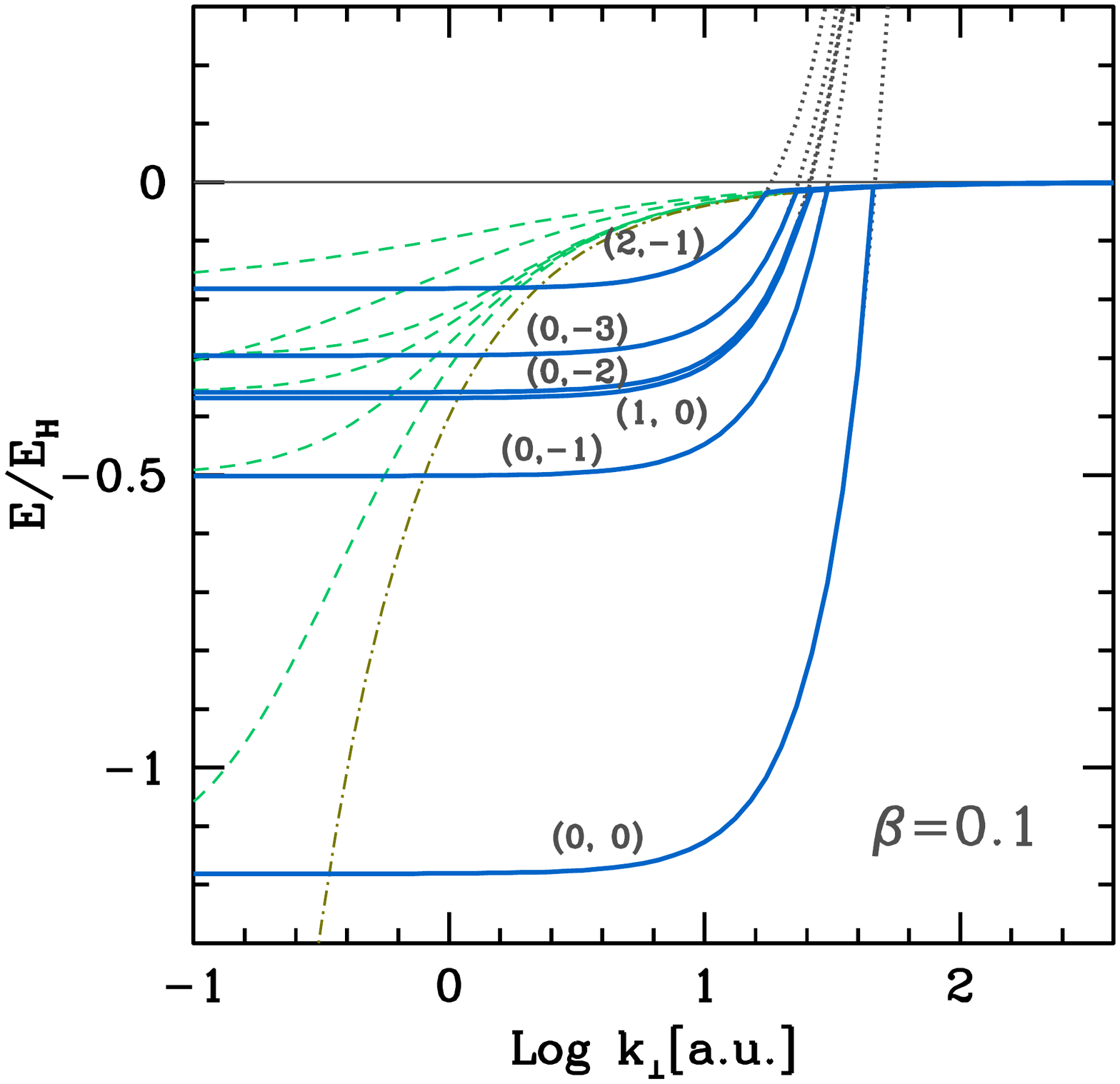}
\caption{Idem Fig. \ref{f:EHk2} but for $\beta=0.1$ ($B=4.70\times 10^{8}$~G).
Long-dashed lines are not shown since they coincide with the dotted lines.
\label{f:EHk1}}
\end{figure}
%=============================================
 
Fig. \ref{f:EHk2} shows with solid lines the energies of some ($\nu,m$) states (the most tightly ones and few others) as a function of the transverse pseudo-momentum and for a strong field ($\beta=500$). For low transverse motions, values grow quadratically with $k_\perp$ according to the perturbative method [Eq. (\ref{E.0})]. Because effective masses $M_\perp$ are larger than $M$, curves remain below the values obtained when the CM effects are ignored (dotted lines). At $k_\perp$ larger than some critical value $\cm{K}_c$ ($\approx 163$, $111$ and $80$ a.u. for the lowest states with $\nu=0$ and $m=0$, $-1$, $-2$, respectively), states become decentered and the energies grow more slowly, according to Eq. (\ref{e.Ev0}). The asymptotic value reached by each state at high transverse motions depends exclusively on $m$. States with $m<0$ rise above the ionization threshold at sufficiently high tranverse motions. In particular, the state $(2,-1)$ is completely embedded in the continuous spectrum on the whole $k_\perp$ domain due to CM effects. In the right extreme of the figure, the curves converge to the approximation given by Eq. (\ref{e.Ek_asymb})  (dot-dashed lines), where the electron is expected to be located around the magnetic well to a distance $r_\text{c}$ from the proton.

It was suggested that general properties about the formation of decentered states remain valid at all field intensities \citep{vincke:1992, baye:1992}.
Fig. \ref{f:EHk1} explores results on the low magnetic field domain, with the application of the present calculations to the case $\beta=0.1$. Two main observations of \citet{vincke:1992} about the extrapolated properties of decentered states at low fields can been appreciated in this figure. First, centered states show low sensitivity to the coupling between internal and global motions. In this branch of the spectrum, the energy values (solid lines) are very similar to standard calculations without CM corrections (dotted lines), because $M_\perp\approx M$. Second, decentered states becomes weakly bound, and fastly converge to a unique curve coincident with the approximation Eq. (\ref{e.Ek_asymb}), i.e., when $k_\perp$ exceeds a critical value, states become abruptly decentered with the electron around the magnetic well.

%%%%%%%%%%%%%%%%%%%%%%%%%%%%%%%%%%%%%%%%%%%%%%%%%%%%%%%%%%%%%%%%%
\section{Ionization equilibrium}\label{s.stat}

The condition for the ionization equilibrium of a gas of magnetized atomic hydrogen ($\text{H}\leftrightarrow \text{H}^++\text{e}^-$) can be written in the usual form based on the chemical potentials $\mu_i$ of the involved species,
\beq \label{e.mu}
\mu_\text{H} = \mu_\text{p}+\mu_\text{e}.
\eeq
The chemical potential expressions adopted in the present work for electrons, protons and atoms are given as follow.

{\it Electrons:}
The energy of a free electron in the field is composed by the Landau levels (including spin-field coupling with giromagnetic ratio $g_\text{e}=2$) plus the kinetic contribution from the motion parallel to the field,
\beq \label{e.Ee}
E = \hbar\omega_\text{e} j +\frac{p_z^2}{2m_\text{e}}, 
\quad j=0,1,2,\dots.
\eeq
The spin degeneracy is one ($\sigma_z=-1$) for the ground Landau level, and it is two ($\sigma_z=\pm1$) for excited levels.
These energies have multiplicity one in $j=0$ and two otherwise. 
From Fermi statistics follows that the electron density in the gas is \citep{potekhin:1999}
\beq \label{e.ned}
n_\text{e} = \frac{\hbar\omega_\text{e}}{\sqrt{\pi}\kB T\lambda_\text{e}^3}
\left[I_{-1/2}\left(\frac{\mu_\text{e}}{\kB T}\right) +2\sum_{j=1}^{\infty}
 I_{-1/2}\left(\frac{\mu_\text{e}}{\kB T}-\frac{\hbar\omega_\text{e} j}{\kB T}\right) \right],
\eeq
with $\kB$ the Boltzmann constant, $T$ the gas temperature, $\mu_\text{e}$ the electron chemical potential, $\lambda_\text{e}=\hbar\sqrt{2\pi/(\kB T m_\text{e})}$  the electron thermal wavelength, and $I_s(x)$ the Fermi integral of order $s$.
In the classical  limit (low density or high temperature so that $-\beta\mu_\text{e}\gg1$, $I_{-1/2}(x)\approx\sqrt{\pi}e^{x}$) Eq. (\ref{e.ned}) reduces to the form given by \citet{gnedin:1974}. In this case, the electron chemical potential is given by
\beq \label{e.mu_e}
\frac{\mu_\text{e}}{\kB T} = 
  \ln\left(\frac{n_\text{e}\lambda_\text{e}^3}{2}\right) 
 +\ln\left[ \left(\frac{2\kB T}{\hbar\omega_\text{e}}\right)\tanh\left(\frac{\hbar\omega_\text{e}}{2\kB T}\right)\right].
\eeq
The last term on the r.h.s. of Eq. (\ref{e.mu_e}) represents the chemical potential excess (respect to the ideal gas contribution) due to the electron interaction with the magnetic field,
\beq \label{e.mu_ex}
\frac{\mu^\text{ex}_\text{e}}{\kB T} 
=\ln\left[\frac{\tanh\left(\eta\right)}{\eta}\right],\hskip.3in
\eeq
which depends on the parameter defined by 
\beq \label{e.x}
\eta\equiv\frac{\hbar\omega_\text{e}}{2\kB T}=\frac{\beta}{\kB T/(2E_\text{H})} .
\eeq
The asymptotic behaviors of $\mu^\text{ex}_\text{e}$ are
\beq
\frac{\mu^\text{ex}_\text{e}}{\kB T} = \left\{ \begin{array}{ll}\displaystyle
-\eta^2/3+ O(\eta^4), & \quad (\eta\ll1),\\
[1ex]\displaystyle -\ln(\eta), & \quad (\eta\gg1).\\
\end{array} \right. 
\eeq

{\it Protons:} The energy of a proton in the field is given by
\beq \label{e.Ep}
E = \hbar\omega_\text{p} j +\frac{p_z^2}{2m_\text{p}}, 
\quad j=0,1,2,\dots.
\eeq
with $\omega_\text{p}=(m_\text{e}/m_\text{p})\omega_\text{e}$. The zero-point and spin-field interaction energies of protons are omitted in (\ref{e.Ep}) because they do not affect the chemical equilibrium. Classical statistics yields
\beq \label{e.mu_p}
\frac{\mu_\text{p}}{\kB T} = 
  \ln\left(n_\text{p}\lambda_\text{p}^3\right) 
 +\ln\left[ \frac{1-\exp\left(-\hbar\omega_\text{p}/\kB T\right)}
  {\hbar\omega_\text{p}/\kB T}\right],
\eeq
where $\lambda_\text{p}=\hbar\sqrt{2\pi/(\kB T m_\text{p})}$ is the proton thermal wavelength, and $n_\text{p}$ the number density of protons. The chemical potential excess of protons is given by
\beq \label{e.mu_px}
\frac{\mu^\text{ex}_\text{p}}{\kB T} = \ln\left[\frac{1-e^{-q\eta}}{q\eta}\right],
\eeq
with $q\equiv 2m_\text{e}/m_\text{p}\approx0.00108923$, and the limits
\beq \label{e.mu_pxa}
\frac{\mu^\text{ex}_\text{p}}{\kB T} = \left\{ \begin{array}{ll}\displaystyle
-q\eta/2+ O(\eta^2), & \quad (\eta\ll1),\\
[1ex]\displaystyle  6.82228-\ln(\eta), & \quad (\eta\gg1),\\
\end{array} \right. 
\eeq

{\it Atoms:} The energy spectrum of bound states of atoms have been detailed in previous section. Energies of centered and decentered states may be written in the general form
\beq \label{e.EE}
E = \cm{E}_\kappa(k_\perp) +\frac{k_z^2}{2M},
\eeq
where $\kappa$ label the set of quantum numbers of a bound state, i.e., $\kappa=\{n,l,m,m_s\}$ at low field and $\kappa=\{N,\nu,m,m_s\}$ for strong field, both connected by the relationships (\ref{e.nu_even})-(\ref{e.n_v}). Besides, the quantity $\cm{E}_\kappa(k_\perp)$ in (\ref{e.EE}) 
is straightfowardly derived from either Eq. (\ref{E.0}) or (\ref{e.Ev0}), taking the minimum value of them. 

The chemical potential of magnetized atoms may be written in the usual form 
\beq \label{e.mu_H}
\mu_\text{H}=\kB T \ln\left(\frac{n_\text{H}\lambda_\text{H}^3}{Z_\text{H}}\right),
\eeq
with $\lambda_\text{H}=\hbar\sqrt{2\pi/(\kB T M)}\approx\lambda_\text{p}$, $n_\text{H}$ the number density of atoms, and $Z_\text{H}$ a partition function which contains non-ideal effects and the coupling of internal and transverse kinetic energies,
\beq \label{e.ZH}
Z_\text{H}= \sum_\kappa\frac{1}{M\kB T}  \int w_\kappa(k_\perp)
   e^{-\cm{E}_\kappa(k_\perp)/(\kB T)}k_\perp dk_\perp ,
\eeq
$w_\kappa(k_\perp)$ being the so-called occupational probability of the state $(\kappa,k_\perp)$. Eqs. (\ref{e.mu_H}) and (\ref{e.ZH}) are derived in the framework of Helmholtz free-energy method \citep{potekhin:1999}. This method is valid at a moderately low density regime, usually $\rho<10^{-2}$~g~cm$^{-3}$ in  field-free conditions \citep{hummer:1988, saumon:1991}, although the well-known reduction of effective sizes of atoms due to the magnetic field may push this limit to a higher value. 
At low temperatures, when most atoms have $k_\perp\ll \cm{K}_c$, the partition function reduces to \citep{pavlov:1993} 
\beq
Z_\text{H}= \sum_\kappa\frac{M_\perp}{M}  w_\kappa
   e^{-\cm{E}_\kappa/(\kB T)},
\eeq
where now $\cm{E}_\kappa$ is the energy of the rest atom. 
Clearly, $Z_\text{H}$ converges to a standard internal partition function when the magnetic field is switch off ($M_\perp\rightarrow M$).

A precise account of particle interactions in the evaluation of species populations at equilibrium, demands the use of quantum-mechanics calculations of collective energies for a reference atom and its nearest neighbor particles, all them under the action of a magnetic field. To the best knowledge of the authors, these calculations are not available.
Since a study of particle interaction effects on the chemical equilibrium is beyond the scope of the present paper, we adopt here a simple scheme \citep{hummer:1988,lai:2001}, which is typically used in model atmospheres of white dwarf stars (e.g., \citet{bergeron:1991, rohrmann:2002}).
\footnote{For a detailed treatment in neutron stars conditions see \citet{potekhin:1999}.}
In this formalism, the occupation probability is given by a product of two contributions
\beq \label{e.occup}
w_\kappa(k_\perp) = w^{(\text{n})}_\kappa \times w^{(\text{c})}_\kappa
\eeq
which express a reduction of the phase space available for the state $\kappa$ as a consequence of statisticaly independent perturbations of the atom due to neutral 
($w^{(\text{n})}_\kappa$) and charged ($w^{(\text{c})}_\kappa$) particles. The neutral contribution is written as
\beq \label{e.occup_neu}
w^{(\text{n})}_\kappa(k_\perp)
      = \exp\left[-n_\text{H} v_\kappa(k_\perp)\right],
\eeq
where $v_\kappa(k_\perp)$ is the effective volume of the atom in the state $(\kappa,k_\perp)$, and it is assumed that bound particles different to atoms (molecules, particle chains, negative ions) are not present in the gas.
Eq. (\ref{e.occup_neu}) results from changes in the gas entropy due to a reduction of the available volume for the atom due to overlapping of electron configurations  in the system. 
On the other hand, for the evaluation of $w^{(\text{c})}_\kappa$ we use results of \citet{nayfonov:1999}. This term can be interpreted as a lowering of the continuum energy level yielded by electric microfield fluctuations originated from neighboring charged particles.

High excited states usually have large atomic size and are very weakly bound, so that Eq. (\ref{e.occup}) gives a low occupation probability for them. Moreover, $w_\kappa$ assures the convergence of the partition function in two ways, yielding a finite number of bound states effectively occupied and an upper value for the allowed values of the pseudo-momentum. In fact, because the limit of atomic sizes ($d$) allowed in a gas with mass density $\rho$ (roughly $\pi d^3/6=M/\rho$), there is a maximum value of pseudo-momentum for decentered states ($d\ga r_c$) 
\beq \label{e.Qtop}
k_\perp^\text{top}\la \frac{2\beta}{a_\text{B}}\left(\frac{6M}{\pi\rho}\right)^{1/3}
  \approx  5.568 \frac{\beta}{\rho[\text{c.g.s.}]^{1/3}}\;\text{a.u.}
\eeq
Accordingly, integration on Eq. (\ref{e.ZH}) is performed with Gaussian quadrature in the range $0\le k_\perp \le 10 k_\perp^\text{top}$ to ensure an appropriate evaluation of $Z_\text{H}$.

%%%%%%%%%%%%%%%%%%%%%%%%%%%%%%%%%%%%%%%%%%%%%%%%%%%%%%%%%%%%%%%%%
\subsection{Effective atomic sizes}\label{s:H_sizes}

The electron cloud (mean probability distribution) in a non-moving atom changes from spherical symmetry at $\beta\ll 1$ to cylindrical one at $\beta\gg 1$. In the zero-field limit, the effective radius of a state $\kappa=(n,l,m)$ is
\beq \label{e.rnl}
r_{nl}=\frac{a_\text{B}}{2}\left[3n^2-l(l+1)\right],
\eeq
with $a_\text{B}$ the Bohr radius. On the other hand, at very intense fields, the root-mean-square of the transversal radius of states $\kappa=(N,\nu,m)$ tends to the Landau state one
\beq\label{e.dperp}
d_\perp = a_\text{L}\sqrt{2n_r+|m|+1} = a_\text{B}\sqrt{\frac{|m|+1}{\beta}},
\eeq
where $a_\text{L}=a_\text{B}/\sqrt{\beta}$ is the Larmor radius and
the last expression in (\ref{e.dperp}) corresponds to bound states ($n_r=0$).

%=============================================
\begin{table}
\caption{Some values of coefficients in Eq.(\ref{e.rd}).\label{t.sizes}}
\setlength{\tabcolsep}{3pt}
\begin{small}
\begin{tabular}{clll}  
\hline 
state & \multicolumn{1}{c}{$b_\perp$} & \multicolumn{1}{c}{$b_z$} & 
\multicolumn{1}{c}{$c_z$} \\ \hline
 1$s_0$    & $0.5$ & $0.2$ & $0.4$ \\ 
 2$s_0$    & $0.8$ & $0.1$ & $0.3$ \\ 
 2$p_0$    & $1.5$ & $0.2$ & $0.3$ \\ 
 2$p_{-1}$ & $2.5$ & $1.2$ & $0.35$ \\ 
\hline
\end{tabular}
\end{small}
\end{table}
%=============================================

An analysis of the spatial probability distribution of the electron for a number of eigenenergy states using data from \citet{ruder:1994} and \citet{schi:2014}, let us find approximated fits for the longitudinal size and transverse radius of the wavefunctions 
\beq \label{e.rd}
d_z = \frac{r_{nl}}{1 + b_z \beta^{c_z}},\hskip.3in
d_\perp = \frac{r_{nl}}{1 + b_\perp \beta^{1/2}},
\eeq
with $b_z$, $b_\perp$ and $c_z$ numerical parameters depending on the state $\kappa$. Some values are given in Table \ref{t.sizes}. 
The effective volume of an atom at rest or moving is approximated by 
\beq
v_\kappa(k_\perp)=\frac{\pi}{6}\left( 2d_\perp^2 + d_z^2 +r_*^2 \right)^{3/2}, 
\eeq
where $r_*=r_*(k_\perp)$ measures the mean electron-proton separation, that is the distance between the mean positions of proton and electron in an atom with transverse pseudo-momentum $k_\perp$. In particular, $r_*\approx 0$ for centered states and  $r_*\rightarrow r_c$ [Eq. (\ref{e.rc})] for strongly decentered states. Analytical expressions can be used to represent $r_*$ following a smooth transition between these limits. However, for numerical calculations  in relatively low densities ($\rho<0.01$~g~cm$^{-3}$) or moderate low field intensities ($\beta\la1$), it is possible to adopt
\beq \label{e.mu_ex}
r_* = \left\{ \begin{array}{ll}\displaystyle
0, & \quad (\text{centered state}),\\
[1ex]\displaystyle  r_\text{c}, & \quad (\text{decentered state}),\\
\end{array} \right. 
\eeq
At the mentioned conditions, we have verified that the occupation probability of decentered states desviates from unity mostly when the approximation (\ref{e.Ek_asymb}) is valid.

%%%%%%%%%%%%%%%%%%%%%%%%%%%%%%%%%%%%%%%%%%%%%%%%%%%%%%%%%%%%%%%%%
\subsection{The law of mass action}\label{s:mass}

According to the condition (\ref{e.mu}) and Eqs. (\ref{e.mu_e}), (\ref{e.mu_p}) and (\ref{e.mu_H}), the equilibrium constant $\cm{Q}$ for the ionization process of hydrogen in a magnetic field is given by
\beq \label{e.Q}
\cm{Q} = \frac{n_\text{H}}{n_\text{e}n_\text{p}}
=\frac{\lambda_e^3}{2} f(\eta) Z_\text{H}^*  e^{\cm{E}_{0}/k_\text{B}T},
\eeq
where
\beq \label{e.Zpavlov}
Z_\text{H}^*=Z_\text{H}e^{-\cm{E}_{0}/k_\text{B}T},
\eeq
$\cm{E}_{0}$ ($>0$) is the ionization energy of magnetized atoms, and the factor $f(\eta)$ comes from the chemical potential excess of electron and proton,
\beq
f(\eta)=\frac{\tanh(\eta)(1-e^{-q\eta})}{q\eta^2}.
\eeq
$Z_\text{H}^*$ represents the partition function defined with the zero-point energy in the ground state ($N=0,\nu=0,m=0,m_s=-\frac12$) and, therefore, is a measure of atomic excitation degree. At low or moderate fields and sufficiently low temperature (far away of pressure ionization conditions), $Z_\text{H}^*$ takes a minimum value between one and two depending of the occupation fraction of the spin-up state at $(\nu,m)=(0,0)$.

Finally, the number density of a state $\kappa$ reads
\beq \label{e.nkappa}
n_\kappa = \frac{n_\text{H}}{M\kB T Z_\text{H}} \int w_\kappa(k_\perp)
   e^{-\cm{E}_\kappa(k_\perp)/(\kB T)}k_\perp dk_\perp .
\eeq
%

%%%%%%%%%%%%%%%%%%%%%%%%%%%%%%%%%%%%%%%%%%%%%%%%%%%%%%%%%%%%%%%%%
\section{Results} \label{s.results}

%=============================================
\begin{figure}
\includegraphics[width=.48\textwidth]{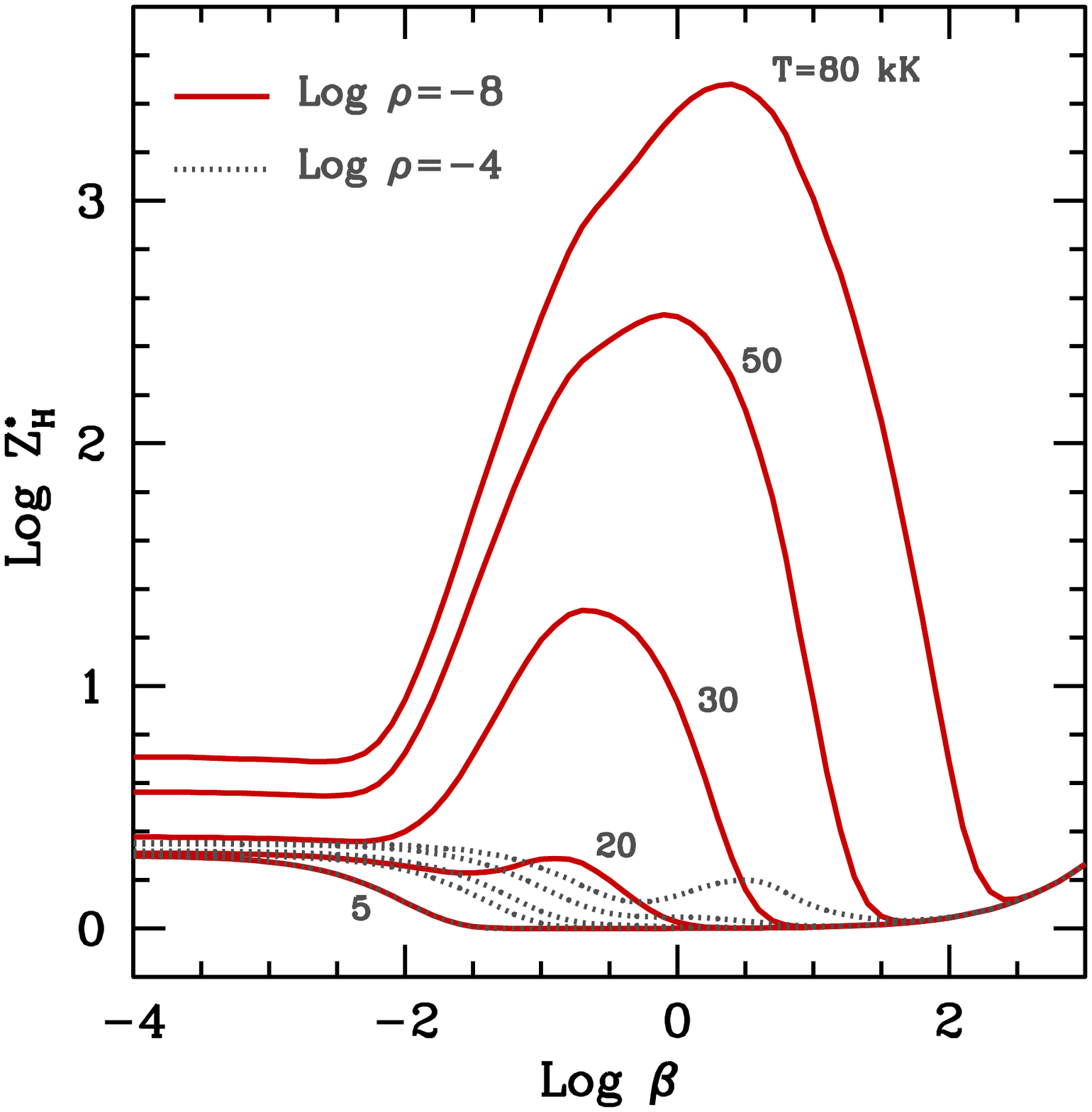}
\caption{Values of the partition function $Z^*_\text{H}$ along isotherms for the two densities indicated on the plot. Temperatures increase from botton to top (labeled for  $\rho=10^{-8}$ g~cm$^{-3}$).
\label{f:z}}
\end{figure}
%=============================================
Current section shows calculations of the equilibrium ionization of magnetized hydrogen carried out for temperatures and densities present in MWD atmospheres.
MWDs constitute about 10\% of the total population of single white dwarfs. As their nonmagnetic analogs, they have thin non-degenerate atmospheres and are mostly composed of hydrogen. Several hundred of MWDs are currently known \citep{kepler:2014}. They have effective temperatures $T_\text{eff}\approx 5000$~--~$80000$~K and gravity surfaces around $\log g\approx8$, whence, the gas density in their atmospheres roughly ranges from 10$^{-12}$ to $10^{-3}$ g~cm$^{-3}$. 

In Fig. \ref{f:z} the run of the partition function versus the magnetic field intensity is shown for several temperatures and densities.
Curves at $T=5000$~K correspond to conditions where the electronic excitations are neglegible in the zero-field limit. For this temperature, the partition function decreases to unity at $\log \beta\la-1.5$ ($B\la10^8$~G) as the energy of the spin-up state in the level $n=1$ shifts toward the continuum. At larger fields, $Z_\text{H}^*$ slowly increases due to contributions of tightly states ($\nu=0$, $m=-1$, $-2$, \dots) coming from high energies (see Fig. \ref{f:EHk0}). Electronic excitations take place for higher temperatures. In particular, isotherms with $T\ge 30000$~K and $\rho=10^{-8}$ g~cm$^{-3}$ (solid lines in the figure) show a strong increasing of $Z_\text{H}^*$ at intermediate field values due to contributions from decentered states. As field rises, partition function curves finaly drop to minimum values (similar to low temperature results) because large energy differences between decentered states and the tightly state ($v,m)=(0,0$), which grow gradually 
with $\beta$. Furthermore, the partition function takes lower values at high densities, as can be appreciate for $\rho=10^{-4}$~g~cm$^{-3}$ (dotted lines in Fig. \ref{f:z}). This is caused by decreased occupation probabilities of excited states, especially decentered states which are destroyed due to their large atomic sizes.

%=============================================
\begin{figure}
\includegraphics[width=.48\textwidth]{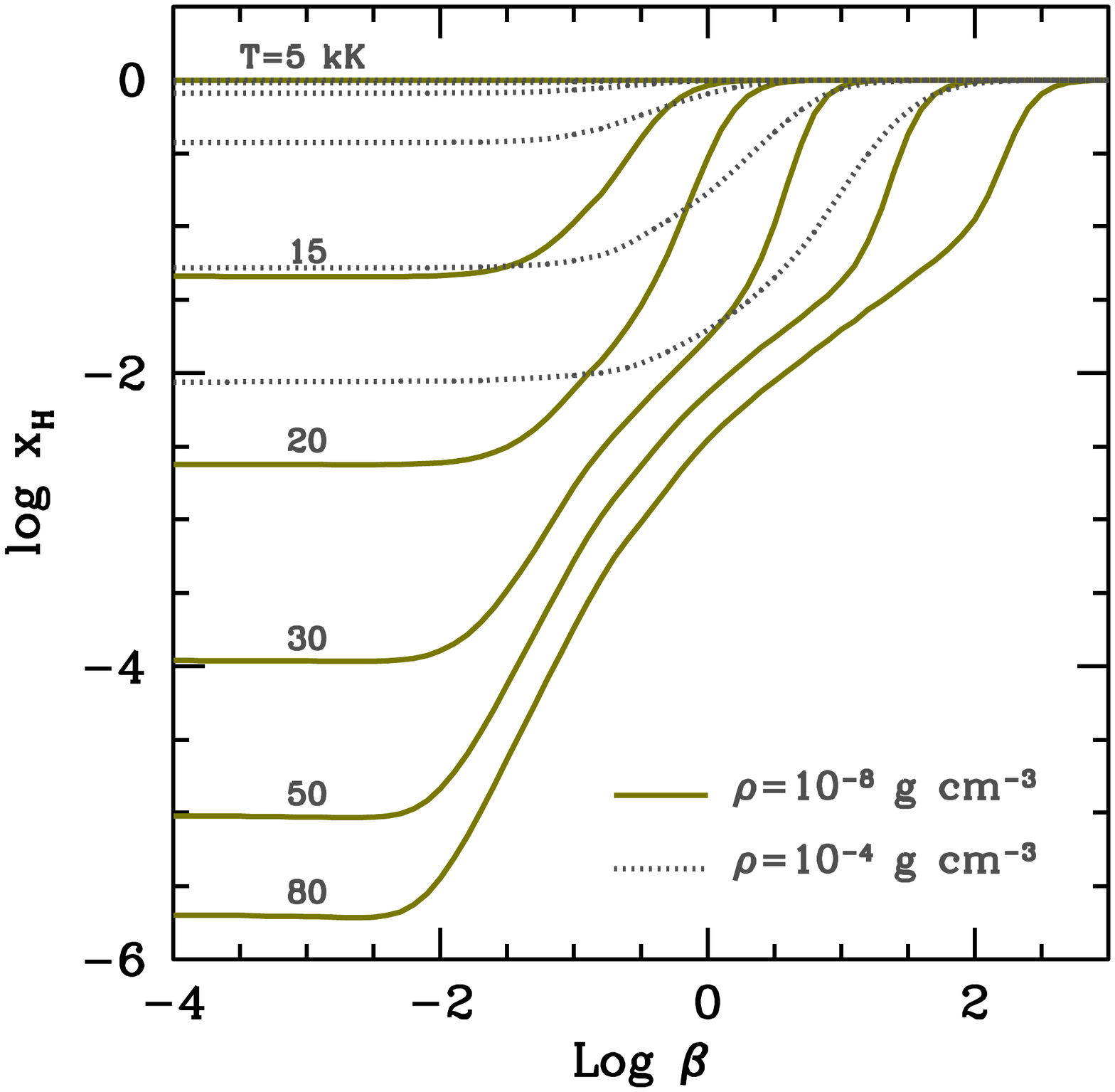}
\caption{Concentration of atoms along isotherms for two densities. Temperatures are indicate on the plot for the case $\rho=10^{-8}$~g~cm$^{-3}$. The same $T$ values correspond to $\rho=10^{-4}$~g~cm$^{-3}$, decreasing from botton to top. 
\label{f:x}}
\end{figure}
%=============================================
Fig. \ref{f:x} shows the neutral atom fraction, $x_\text{H}=n_\text{H}/(n_\text{H}+n_\text{p})$, derived from the equilibrium condition (\ref{e.Q}) and the barionic conservation equation assumed in the present work ($\rho/M=n_\text{H}+n_\text{p}$). Magnetic effects on the equilibrium constant coming from  free electrons and protons are represented by the factor $f(\eta)$, which decreases monotonically from one to zero as the field intensity grows. Therefore, $f(\eta)$ favors the gas ionization. However its effect is usually countared by other contributions to $\cm{Q}$, particularly by $Z_\text{H}^*$ at moderate fields and by the Boltzmann factor (containg the ionization energy $\cm{E}_0$) at intense fields. In fact, as the field strength is raised over $\beta\approx0.01$, the Boltzmann factor  strongly increases (through the ionization energy) and ultimately overcomes the equilibrium constant. As a consequence, the gas tends to a full recombination at high magnetic fields. Abundance of atoms grows as the density increases to $\rho\approx10^{-4}$~g~cm$^{-3}$ (dotted lines in Fig. \ref{f:x}), while the ground state and tightly states remain unperturbed (occupation probability close to unity).

%=============================================
\begin{figure}
\includegraphics[width=.48\textwidth]{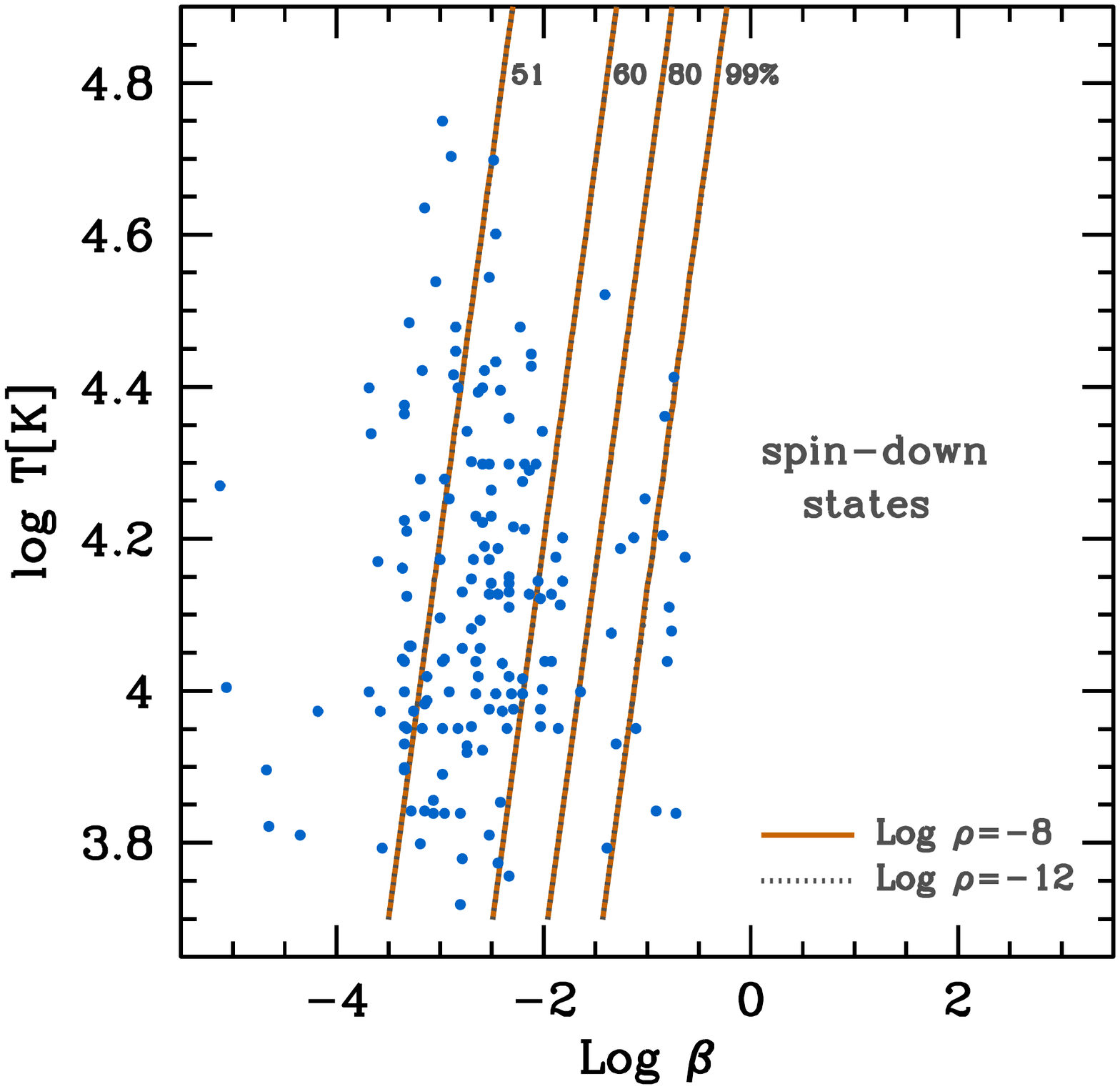}
\caption{Contributions in percent of spin-down states to the partition function $Z_\text{H}^*$. Results are shown for two densities, $\rho=10^{-8}$ and $10^{-12}$ ~g~cm$^{-3}$. No sensitive dependence on the density is appreciate. 
Circles represent the mean conditions (effective temperature versus field intensity) found in the atmospheres of a sample of known MWDs \citep{kulebi:2009}.  
\label{f:zspin}}
\end{figure}
%=============================================

%=============================================
\begin{figure}
\includegraphics[width=.48\textwidth]{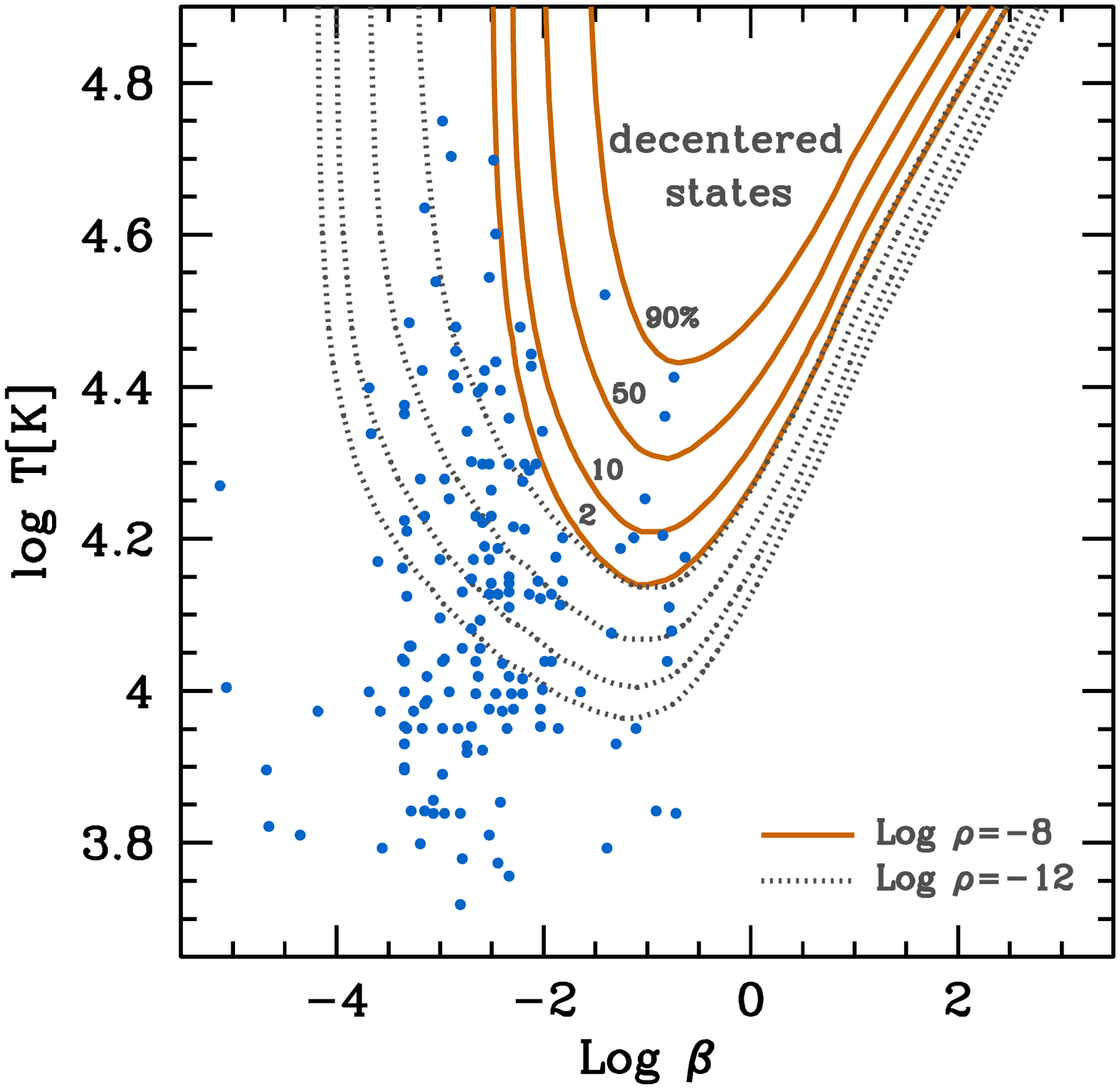}
\caption{Same as Fig. \ref{f:zspin}, but for the fraction (in percent) of partition function $Z_\text{H}^*$ coming from decentered states. Values are indicated on the plot for $\rho=10^{-8}$ g~cm$^{-3}$.
\label{f:zdec}}
\end{figure}
%=============================================
Various effects of the magnetic field on the gas in the conditions of atmospheres of MWDs may be explored. On the one hand, Fig. \ref{f:zspin} illustrates on a plane of temperature versus magnetic strength (in logarithmic scales) the mean values determined for a sample of MWDs \citep{kulebi:2009}, representing physical conditions in their atmospheres. Lines show percentages of the spin-down state contribution in the partition function. It can be clearly seen in the figure that spin-up states are removed of the partition function for $\log \beta>-3$ just over the bulk of strong MWDs. This result is closely independent on the density in the regime studied.

On the other hand, Fig. \ref{f:zdec} shows the importance of decentered states relative to centered ones on the partition function $Z_\text{H}^*$. Decentered states becomes dominant at high temperatures and intermediate values of the magnetic field, with peaks around $\beta\approx 0.1$, just where the strongest MWDs are located. The contribution of decentered states extend to lower temperatures and a wide range of field values for low densities ($\rho = 10^{-12}$ ~g~cm$^{-3}$ in the figure). According to these results, decentered states could be present in the outer layers of hot MWDs ($T_\text{eff}\ga20000$~K) with megagauss fields and in deep atmospheric layers of the strongest magnetic objects. Present calculations are based on extrapolations  towards low magnetic fields of results of CM effects derived at higher fields, following the expected by \citet{vincke:1992} and \citet{baye:1992}. It is clear that valuable insight can be gained if new accurate studies are aimed to analyze the coupling between the internal energy of atoms and their transverse movements at moderate and weak magnetic fields. These studies will be useful to search and identify possible spectroscopic signatures of decentered states in the spectrum of MWDs.

%=============================================
\begin{figure}
\includegraphics[width=.48\textwidth]{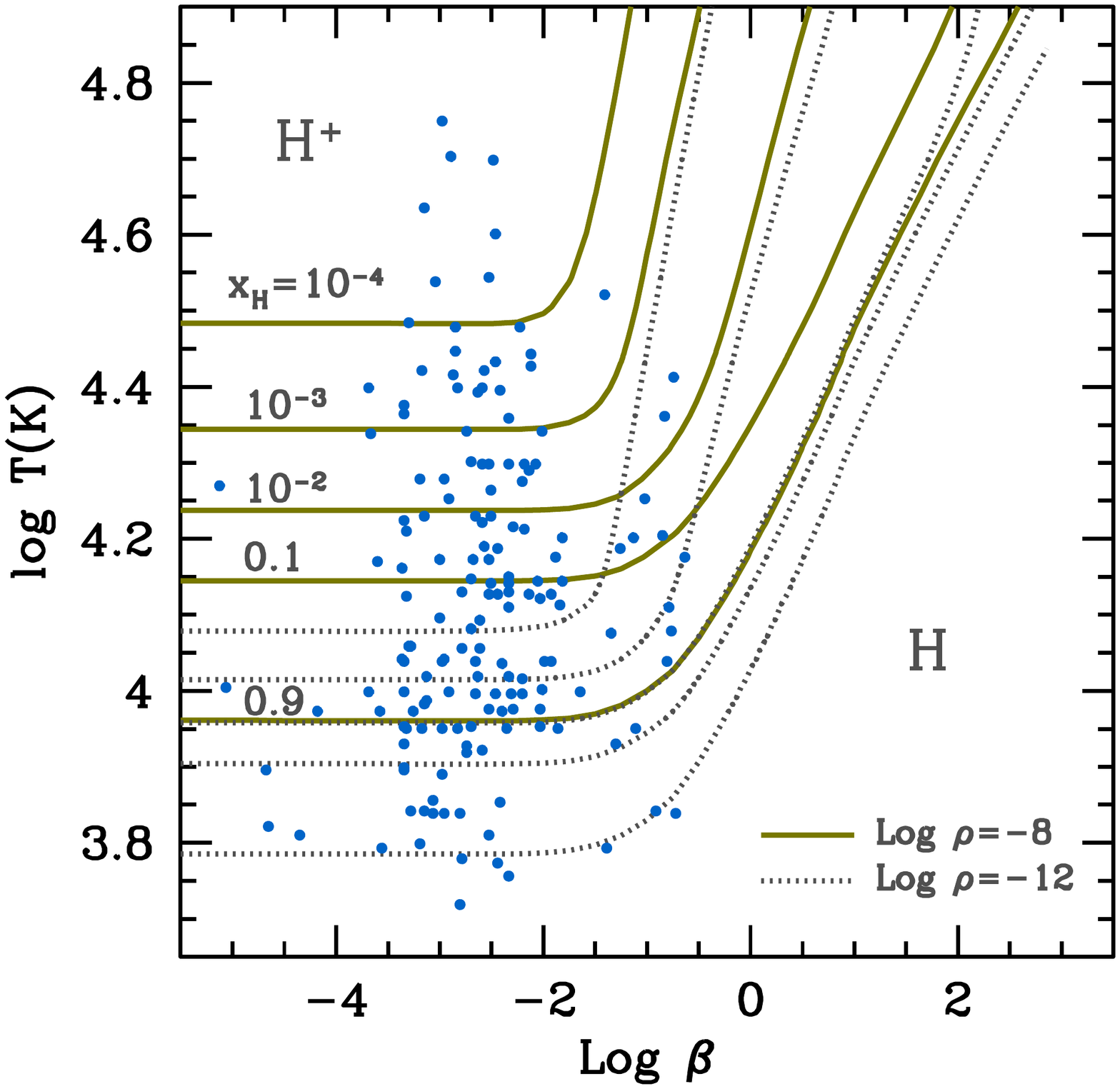}
\caption{Same as Fig. \ref{f:zspin}, but for concentrations of neutral atoms in the gas. Values of $x_H$ are indicated on the plot for $\rho=10^{-8}$ g~cm$^{-3}$.
\label{f:xgral}}
\end{figure}
%=============================================
Fig. \ref{f:xgral} displays the predicted curves of iso-abundance of atoms in the usual diagram of temperature against field intensity. It is inmediately clear that the abundance of neutral atoms does not stay constant along the sample of known MWDs. The onset of magnetic recombination occurs approximadely at $\log\beta=-2$ ($B\approx 5\times 10^7$~G), decreasing the gas ionization for the strongest MWDs observed. The electronic recombination increases with the gas density in the whole density regime of these atmospheres, since lowest energy states remain barely perturbed by particle neighbors at $\rho\la10^{-3}$~g~cm$^{-3}$. It is worth noting that changes in the ionization equilibrium can affect the structure of an atmosphere, specially througth its influence on the radiative opacities.

%%%%%%%%%%%%%%%%%%%%%%%%%%%%%%%%%%%%%%%%%%%%%%%%%%%%%%%%%%%%%%%%%
\section{Final remarks} \label{s.concl}

We have given a general method for calculating the partition function of hydrogen atoms in a wide range of field strengths. The evaluations rely on fits of accurate energy evaluations of atoms at rest, complemented by CM effects on internal atomic structure, particularly, those arised from thermal motions across the magnetic field. 
The work is benefied by mathematical relations which express the correspondence between sets of quantum numbers at the limits of zero-field and very strong fields. These relations are given here for the first time.

To our knowledge, this is also the first time that a detailed treatment of the atomic hydrogen abundance under ionization process is undertaken for magnetic fields from zero to very high strengths.
In particular, we extended the analysis of ionization balance of hydrogen to include the regime of moderate magnetic fields ($\beta\la1$, $B\la 10^{9}$~G). This compresses the realm of magnetic white dwarfs, where the effects of the field on the chemical equilibrium are not included in current atmosphere models. We found however that these effects can not be ignored at least for strongest MWDs.
In fact, although such fields are much lower than those achieved in neutron stars, they are sufficiently strong to modify the ionization equilibrium. 
Future work will be devoted to include the formation of molecules and other species (which are also altered by the magnetic field) in order to obtain reliable occupation numbers of atomic states and free particles.

It was shown here that decentered states could be present in the surface of strong MWDs, being the dominant contribution of the partition function in some cases (hot and strongly magnetic atmospheres). If this phenomenon occurs, it can have effects on the energy distribution emitted by these stars. 
A definitive analysis of decentered state role on MWDs requires further investigation on the formation of these states at weak magnetic fields and, in general, on the relationship between the movement of atoms across the field and their internal energy spectrum.

\begin{acknowledgements}
We wish to thank C. Schimeczek, who kindly sent us his energy evaluations for magnetized atoms at rest, and Jos\'e Ignacio Castro for useful discussions. Authors thank the support of the MINCYT (Argentina) through Grant No. PICT 2016-1128. 
\end{acknowledgements}

%%%%%%%%%%%%%%%%%%%%%%%%%%%%%%%%%%%%%%%%%%%%%%%%%%%%%%%%%%%%%%%%%
\appendix

%%%%%%%%%%%%%%%%%%%%%%%%%%%%%%%%%%%%%%%%%%%%%%%%%%%%%%%%%%%%%%%%%
\section{Analytical fits} \label{a:1}

We show analytic fits to energy data of \citet{schi:2014} for bound states of hydrogen atoms at rest in the approximation of infinity nuclear mass. Results are expressed in logarithmic scales. In this appendix we adopt the notation ($\log\equiv\log_{10}$)
\beq \label{eA.a}
x=\log\beta, 
\eeq
\beq \label{eA.b}
\epsilon=\log(-\cm{E}_\infty/E_\text{H}),
\eeq
\beq
\epsilon_n=\log(-E_{n}/E_\text{H})=-2\log(n),
\eeq
\beq 
\epsilon_\nu=-2\log\left[\text{Int}\left(\frac{\nu+1}2\right)\right].
\eeq
Fits were performed in the range $-4\le \log\beta\le 3$, although they are well behaved in the full non-relativistic domain ($\log\beta<4$)
\footnote{It is worth to recall that relativity effects remain negligible even at much larger values of $\beta$. Accurate numerical solution shows that relativistic corrections to binding energies do not exceed $0.01E_\text{H}$ as long as $\log\beta<6$ \citep{nakashima:2010}.}.
In particular, they converge to the right limits at $\beta=0$ and $\log\beta\gg1$. Fitting equations are organized by the quantum number $\nu$.

%%%%%%%%%%%%%%%%%%%%%%%%%%%%%%%%
\subsection{States $\nu=0$} \label{s.nu0}

Energies of tighly bound states ($\nu=0$, $m=0,-1,-2,\dots$) are represented by
\beq \label{eA.v0}
\epsilon = \left\{ \begin{array}{ll}\displaystyle
1+\frac{\epsilon_n-1}{1+a_1 \exp\{a_2[x-x_a-0.1(x-x_a)^2]\}}, & (x<x_a)\\
[3ex]\displaystyle 
\epsilon_a + (\epsilon_b-\epsilon_a)\left(\frac{x-x_a}{x_b-x_a}\right)^{1.22}, 
& (x_a<x<x_b)\\
[3ex]\displaystyle 
\epsilon_b +(\epsilon_c-\epsilon_b)\left(\frac{x-x_b}{3-x_b}\right)^{0.92},
& (x_b<x)\\
\end{array} \right.
\eeq
with
\beq 
a_1=\frac{\epsilon_n-\epsilon_a}{\epsilon_a-1},
\eeq
\beq
a_2=-\epsilon_a'\frac{(1+a_1)^2}{(\epsilon_n-1)a_1}.
\eeq
The $m$-dependence of parameters ($x_a$, $x_b$, $\epsilon_a$, $\epsilon_b$, $\epsilon_c$, $\epsilon'_a$) is very well adjusted by
\beq\label{fln}
f=b_0+b_1 [\log(1+|m|)]^{b_2}.
\eeq
Values of coefficients $b_j$ are listed in Table \ref{T.0}.
Eq.(\ref{eA.v0}) fits results of \citet{schi:2014} with relative error within few percents (see Fig. \ref{f:v0}).

%=============================================
\begin{table}
\caption{Values of coefficients of quantities calculated with Eq.(\ref{fln}). Numbers in brackets indicate power of ten.\label{T.0}}
\setlength{\tabcolsep}{3pt}
\begin{small}
\begin{tabular}{clll}  
\hline 
Quantity & \multicolumn{1}{c}{$b_0$} & \multicolumn{1}{c}{$b_1$} & 
\multicolumn{1}{c}{$b_2$} \\ \hline
 &  & ($\nu=0$) & \\ \hline
 $x_a$        & $-8.51584(-1)$ & $-2.90213$     & $1.01555$ \\
 $x_b$        & $+7.86224(-1)$ & $-2.28335$     & $9.37692(-1)$ \\
 $\epsilon_a$ & $+9.50091(-2)$ & $-1.97412$     & $1.00523$ \\
 $\epsilon_b$ & $+5.73409(-1)$ & $-1.54066$     & $9.77581(-1)$ \\
 $\epsilon_c$ & $+1.26974$     & $-3.78015(-1)$ & $9.10852(-1)$ \\
 $\epsilon'_a$& $+1.70505(-1)$ & $+5.16550(-2)$ & $6.92991(-1)$ \\
\hline 
 &  & ($\nu=1$) & \\ \hline
 $x_b$        & $-6.66302(-1)$ & $-1.50237$     & $1.17845$ \\
 $\epsilon_b$ & $-3.61037(-1)$ & $-9.80935(-1)$ & $1.22078$ \\
 $\epsilon'_b$& $+2.13743(-1)$ & $+2.23000(-1)$ & $8.82388(-1)$ \\
\hline 
 &  & ($\nu=2$) & \\ \hline
 $x_b$        & $+5.28777(-2)$ & $-2.38204$     & $9.60364(-1)$ \\
 $\epsilon_b$ & $-4.52254(-1)$ & $-1.00281$     & $1.23880$ \\
 $\epsilon'_b$& $+1.19340(-1)$ & $+2.96234(-1)$ & $1.03199$ \\
\hline 
 &  & ($\nu=3$) & \\ \hline
 $x_b$        & $-7.10984(-1)$ & $-1.78597$     & $1.16795$ \\
 $\epsilon_b$ & $-7.90709(-1)$ & $-7.84790(-1)$ & $1.36181$ \\
 $\epsilon'_b$& $+1.17903(-1)$ & $+2.60062(-1)$ & $1.11196$ \\
\hline
\end{tabular}
\end{small}
%\tablefoot{...}
\end{table}
%=============================================

%%%%%%%%%%%%%%%%%%%%%%%%%%%%%%%%
\subsection{States $\nu=1$} \label{s.nu1}

Energies of states with $\nu=1$ are given by a global approximation
\beq\label{eA.v1}
\epsilon= \epsilon_\nu + \frac{\epsilon_n - \epsilon_\nu}
   {1+a_1\exp\{a_2(x-x_b)[1-\delta(x-x_b)]\}},
\eeq
\beq 
a_1=\frac{\epsilon_n-\epsilon_b}{\epsilon_b-\epsilon_\nu},
\eeq
\beq
a_2=-\epsilon_b'\frac{(1+a_1)^2}{(\epsilon_n-1)a_1},
\eeq
with $n=|m|+2$, $\delta =0.26$ except at $|m|=1$, $2$, $3$ (where $\delta=0.20$, $0.22$, $0.24$, respectively) and $x>x_b$ ($\delta=0$). Quantities $x_b$, $\epsilon_b$ and $\epsilon_b'$ are calculated with help of (\ref{fln}) and Table \ref{T.0}. The performance of fits at $\nu=1$ is shown in Fig. \ref{f:v1}.
%

%%%%%%%%%%%%%%%%%%%%%%%%%%%%%%%%
\subsection{States $\nu=2$} \label{s.nu2}

Energies of states corresponding to $\nu=2$ are represented by a piecewise approximation. For $x\le x_b$, we use Eq. (\ref{eA.v1}) with $n=|m|+2$, $\delta=0.2$, supplemented by Eq.(\ref{fln}) with data in Table \ref{T.0}.
For $x>x_b$, $\epsilon$ is given by
\beq
\epsilon= \epsilon_b + \frac{2}{\pi}(\epsilon_b-\epsilon_{n=1})
\tan^{-1}\left[\frac{\pi\epsilon'_b(x-x_b)}{2(\epsilon_b-\epsilon_{n=1})}
[1+\xi(x-x_b)], \right],
\eeq
with
\beq
\xi = -0.0125 +0.030456[\log(1+m_*)]^{1.134}
\eeq
and $m_*=\min(|m|,4)$.

%%%%%%%%%%%%%%%%%%%%%%%%%%%%%%%%%%%%%%%%%%%%%%%%%%%%%%%%%%%%%%%%%
\subsection{States $\nu=3$} \label{s.nu3}

We adopt a piecewise approximation for energies at $\nu=3$. As before, Eq. (\ref{eA.v1}) is applied for $x\le x_b$, with $n=|m|+3$ and $\delta=0$, $0.08$, $0.13$, $0.15$, $0.165$ and $0.17$ for $|m|=0$, 1, 2, 3, 4, 5, and $\ge 6$, respectively. Parameters are represented by Eq.(\ref{fln}) with data in Table \ref{T.0}. For $x>x_b$, the following expression is used
\beq
\epsilon= \epsilon_b + \frac{(\epsilon_\nu-\epsilon_b)(x-x_b)}
{ \{[(\epsilon_\nu-\epsilon_b)/\epsilon'_b]^q+(x-x_b)^q\}^{1/q} },
\eeq
with $q=2.5$.

%%%%%%%%%%%%%%%%%%%%%%%%%%%%%%%%%%%%%%%%%%%%%%%%%%%%%%%%%%%%%%%%%
\subsection{States $\nu\ge4$} \label{s.nu4}

Energy of states with $\nu\ge4$ are calculated as follows
\beq \label{eA.vv}
\epsilon = \left\{ \begin{array}{ll}\displaystyle
\epsilon_a + \frac{(\epsilon_a-\epsilon_{n})(x-x_a)}
{\left[c_a^2+(x-x_a)^2\right]^{1/2}}, & (x\le x_a)\\
[3ex]\displaystyle 
 \epsilon_a + (\epsilon_*-\epsilon_a)\left[
\frac{(\epsilon_b-\epsilon_*) \zeta + (\epsilon_b-\epsilon_a) (x-x_*)^2}
{c_*+(\epsilon_b-\epsilon_a)^2 (x-x_*)^2}
\right], & (x_a,x_b)\\
[3ex]\displaystyle 
 \epsilon_b + \frac{(\epsilon_\nu-\epsilon_b)(x-x_b)}
{\left[c_b^2+(x-x_b)^2\right]^{1/2}}, & (x_b\le x)\\
\end{array} \right.
\eeq
with 
\beq
c_a = \frac{\epsilon_a-\epsilon_{n}}{y_a},\quad
c_*=(\epsilon_*-\epsilon_a)(\epsilon_b-\epsilon_*)\zeta^2,\quad
c_b = \frac{\epsilon_\nu-\epsilon_b}{y_b},
\eeq
\beq
\zeta=x_b-x_a,\quad
\epsilon_*=\frac{\epsilon_b-\epsilon_a}{2},\quad
x_* = \frac{x_a(\epsilon_*-\epsilon_a) + x_b(\epsilon_b-\epsilon_*)}
{\epsilon_b-\epsilon_a}.
\eeq
Besides, $x_a$, $x_b$, $\epsilon_a$, $\epsilon_b$, $y_a$, $y_b$ are fitted by Eq. (\ref{fln}). In turn, parameters $b_j$ ($j=0,1,2$) in this equation are calculated as follows
%
%       ======   b_0   ======
%
\beq \label{e4.b0a}
b_0[x_a] = \left\{ \begin{array}{ll}\displaystyle
-1.1 \tau -1.154902 -2.087178 \Delta^{1.082710}, & (\text{even}~\nu)\\
[1ex]\displaystyle 
-\tau -1.18 -2.312886\Delta^{0.7737455}, & (\text{odd}~\nu)\\
\end{array} \right.
\eeq
\beq
b_0[x_b] = \left\{ \begin{array}{ll}\displaystyle
-0.522879, & (\text{even}~\nu)\\
[1ex]\displaystyle 
-1.154902, & (\text{odd}~\nu)\\
\end{array} \right.
\eeq
\beq
b_0[\epsilon_a] = \left\{ \begin{array}{ll}\displaystyle
-0.68 -1.176143\Delta^{0.8685913}, & (\text{even}~\nu)\\
[1ex]\displaystyle 
-0.9558838 -1.069160\Delta^{0.8065575}, & (\text{odd}~\nu)\\
\end{array} \right.
\eeq
\beq
b_0[\epsilon_b] = \left\{ \begin{array}{ll}\displaystyle
-0.8867395 -1.744739\Delta^{1.095173}, & (\text{even}~\nu)\\
[1ex]\displaystyle 
-1.12 -1.707775\Delta^{1.119483}, & (\text{odd}~\nu)\\
\end{array} \right.
\eeq
\beq
b_0[y_a] = \left\{ \begin{array}{ll}\displaystyle
0.02 -0.034\tau +\frac{0.2}{\nu^{1.1} + 3}, & (\text{even}~\nu)\\
[2ex]\displaystyle 
0.013-0.034\tau +\frac{0.2}{\nu^{0.74} +6}, & (\text{odd}~\nu)\\
\end{array} \right.
\eeq
\beq
b_0[y_b] = \left\{ \begin{array}{ll}\displaystyle
0.3780437\nu^{-0.9572978}, & (\text{even}~\nu)\\
[2ex]\displaystyle 
0.3480917\nu^{-0.9508739}, & (\text{odd}~\nu)\\
\end{array} \right.
\eeq
%
%       ======   b_1   ======
%
\beq
b_1[x_a] = \left\{ \begin{array}{ll}\displaystyle
 -0.01890508, & (\text{even}~\nu)\\
[1ex]\displaystyle 
  0.1044253, & (\text{odd}~\nu)\\
\end{array} \right.
\eeq
\beq
b_1[x_b] = \left\{ \begin{array}{ll}\displaystyle
-0.95 -1.1\nu^{-0.4}, & (\text{even}~\nu)\\
[1ex]\displaystyle 
-0.1 -2\nu^{-0.4}, & (\text{odd}~\nu)\\
\end{array} \right.
\eeq
\beq
b_1[\epsilon_a] = \left\{ \begin{array}{ll}\displaystyle
-0.2 -1.1\nu^{-0.4}, & (\text{even}~\nu)\\
[1ex]\displaystyle 
-0.2 -0.9\nu^{-0.4}, & (\text{odd}~\nu)\\
\end{array} \right.
\eeq
\beq
b_1[\epsilon_b] = \left\{ \begin{array}{ll}\displaystyle
-2.487767\nu^{-0.9652760}, & (\text{even}~\nu)\\
[1ex]\displaystyle 
-2.713701\nu^{-1.000845}, & (\text{odd}~\nu)\\
\end{array} \right.
\eeq
\beq
b_1[y_a] = \left\{ \begin{array}{ll}\displaystyle
0.1438085, & (\text{even}~\nu)\\
[1ex]\displaystyle 
0.1457385, & (\text{odd}~\nu)\\
\end{array} \right.
\eeq
\beq
b_1[y_b] = \left\{ \begin{array}{ll}\displaystyle
0.8265754\nu^{-0.9347425}, & (\text{even}~\nu)\\
[1ex]\displaystyle 
1.0286911\nu^{-0.9818393}, & (\text{odd}~\nu)\\
\end{array} \right.
\eeq
\beq
b_2[x_a] = \left\{ \begin{array}{ll}\displaystyle
0.5904491, & (\text{even}~\nu)\\
[1ex]\displaystyle 
0.7094884, & (\text{odd}~\nu)\\
\end{array} \right.
\eeq
\beq
b_2[x_b] = \left\{ \begin{array}{ll}\displaystyle
0.85 +1.1\nu^{-0.4}, & (\text{even}~\nu)\\
[1ex]\displaystyle 
0.90 +1.5\nu^{-0.4}, & (\text{odd}~\nu)\\
\end{array} \right.
\eeq
\beq
b_2[\epsilon_a] = \left\{ \begin{array}{ll}\displaystyle
0.6 +0.8\Delta^{0.4}, & (\text{even}~\nu)\\
[1ex]\displaystyle 
0.7 +0.73\Delta^{0.4}, & (\text{odd}~\nu)\\
\end{array} \right.
\eeq
\beq
b_2[\epsilon_b] = \left\{ \begin{array}{ll}\displaystyle
1.209001, & (\text{even}~\nu)\\
[1ex]\displaystyle 
1.251972, & (\text{odd}~\nu)\\
\end{array} \right.
\eeq
\beq
b_2[y_a] = \left\{ \begin{array}{ll}\displaystyle
1.596943, & (\text{even}~\nu)\\
[1ex]\displaystyle 
1.603306, & (\text{odd}~\nu)\\
\end{array} \right.
\eeq
\beq
b_2[y_b] = \left\{ \begin{array}{ll}\displaystyle
1.114659, & (\text{even}~\nu)\\
[1ex]\displaystyle 
1.181565, & (\text{odd}~\nu)\\
\end{array} \right.
\eeq
where
\beq \label{e4.D}
\Delta = \left\{ \begin{array}{ll}\displaystyle
\log(\nu)-\log(4), & (\text{even}~\nu)\\
[3ex]\displaystyle 
\log(\nu)-\log(5), & (\text{odd}~\nu)\\
\end{array} \right.
\eeq
\beq \label{e4.tau}
\tau=t-\text{Int}[t]
\eeq
and
\beq \label{e4.t}
t = 2\sqrt{\left[\frac12\left(\nu+\frac12\right)\right]}\pm 2, 
\eeq
with sign $+$ ($-$) for even (odd) $\nu$.

In order to avoid crossing levels, when current expressions give $x_a>x_b$, the corresponding mean values $(x_a+x_b)/2$ and $(\epsilon_a+\epsilon_b)/2$ are assigned for $x_a=x_b$ and $\epsilon_a=\epsilon_b$, respectively.

%%%%%%%%%%%%%%%%%%%%%%%%%%%%%%%%
\subsection{Effective mass of transverse motions} \label{s.mass}

The effective mass given by Eqs. (\ref{e.Mtrav}) and (\ref{e.delta}) may be approximated by
\beq
M_\perp = M\left(1+b\beta^c\right).
\eeq
With energy values calculated by \citet{schi:2014}, we found for $m=0$
\beq
b= \left\{ \begin{array}{ll}\displaystyle
1.78\times10^{-3}(1+\nu)^{2.58}, & \quad (\text{even}~\nu),\\
[1ex]\displaystyle 
10^{-3}(2+\nu)^{3}, & \quad (\text{odd}~\nu).\\
\end{array} \right. 
\eeq
\beq
c= \left\{ \begin{array}{ll}\displaystyle
1+0.27\log(\nu+0.4), & \quad (\text{even}~\nu),\\
[1ex]\displaystyle 
1.02+0.3\log(v+0.4), & \quad (\text{odd}~\nu).\\
\end{array} \right. 
\eeq
%

%%%%%%%%%%%%%%%%%%%%%%%%%%%%%%%%%%%%%%%%%%%%%%%%%%%%%%%%%%%%%%%%%
\bibliographystyle{aa}
\bibliography{biblio_H}
\end{document}